\documentclass[12pt,preprint]{aastex}

\slugcomment{Accepted  for publication in 
 {\it the Astrophysical Journal} on 13th of April 2012}  
\usepackage{natbib}

\def\lax {\ifmmode{_<\atop^{\sim}}\else{${_<\atop^{\sim}}$}\fi}
\def\gax {\ifmmode{_>\atop^{\sim}}\else{${_>\atop^{\sim}}$}\fi}
\def\gtorder{\mathrel{\raise.3ex\hbox{$>$}\mkern-14mu
             \lower0.6ex\hbox{$\sim$}}}

\def\Ne{N_\mathrm{e}}
\def\me{m_\mathrm{e}}
\def\kte{kT_\mathrm{e}}

\def\ktbb{kT_\mathrm{bb}}
\def\taueff{\tau_\mathrm{eff}}
\def\Ec{E_\mathrm{c}}
\def\rph{R_\mathrm{ph}}
\def\Ep{E_\mathrm{p}}

\def\me{m_\mathrm{e}}

\def\chiq{$\chi^2$}
\def\s1{s$^{-1}$}
\def\cm2{cm$^{-2}$}
\def\ster1{ster$^{-1}$}
\def\Vb{V_\mathrm{b}}

\def\fb{f_{\mathrm b}}

\def\xspec{{\sc xspec}}
\def\bmc{{\sc bmc}}
\def\fermi{{\it Fermi}}
\def\sax{{\it BeppoSAX}}
\def\swift{{\it Swift}}

\shorttitle{An upscattering  model for the prompt emission of GRB}
\shortauthors{Titarchuk et al.}

\begin{document}

\title{An upscattering spectral formation model for the prompt emission of Gamma-Ray Bursts}

\author{ Lev Titarchuk\altaffilmark{1,2, 4}, Ruben Farinelli\altaffilmark{1},
 Filippo Frontera\altaffilmark{1} and Lorenzo Amati\altaffilmark{3}, 
}
\altaffiltext{1}{Dipartimento di Fisica, Universit\`a di Ferrara, via Saragat 1, 44122 Ferrara, Italy, email:titarchuk@fe.infn.it}
\altaffiltext{2}{George Mason University, Fairfax, VA 22030, USA, email: ltitarch@gmu.edu}
\altaffiltext{3}{INAF-IASF, Sezione di Bologna, via Gobetti 101, 40129 Bologna, Italy}
\altaffiltext{4}{NASA/Goddard Space Center, Greenbelt, 20770,  USA, email: lev@milkyway.gsfc.nasa.gov}

\begin{abstract}

We propose a model for the spectral formation of Gamma Ray Burst (GRB) prompt emission, 
where the phenomenological Band's function is usually  applied to describe the GRB prompt emission.
We suggest that  the GRB prompt emission is mainly  a result 
{of two upscattering processes. The first process is the Comptonization of  
relatively cold  soft photons of the star off  electrons of a hot shell  of plasma  of  
temperature $T_e$  of the order of $10^{9}$ K (or $\kte\sim 100$ keV) that  moves 
sub-relativistically with the bulk velocity  $V_{\rm b}$ substantially less than 
the speed of light $c$.}
In this phase,  the Comptonization parameter  $Y$ 
is  high  and the interaction between a blackbody-like soft seed photon population and hot 
electrons leads to formation of a saturated Comptonization spectrum 
modified by the sub-relativistic  bulk outflow. {The second process is an upscattering of the
previously Comptonized spectrum by the plasma outflow once it becomes relativistic. This 
process gives rise to  the high-energy power-law component above the peak in the $EF(E)-$diagram where $F(E)$} 
is the energy flux.
The latter process can be described  by a convolution of the Comptonized spectrum  with  a 
broken-power-law Green function. {Possible physical scenarios for this second upscattering
process} are discussed.
{In the framework of our model, we give an interpretation of
the Amati relation between the intrinsic spectral peak photon energy and radiated energy or 
luminosity, and we propose a possible explanation of the GRB temporal variability.}

\end{abstract} 

\keywords{radiation mechanisms: general --- radiative transfer --- gamma-ray burst: general --- stars: massive}

\section{Introduction}

Understanding the 
physical processes which give rise to the observed spectra of the prompt emission of Gamma Ray  Bursts (GRB) is presently one of the most exciting issues studied by both the theoretical and observational community.
The Band function \citep{band93} up to now  widely used to describe their prompt emission is a pure phenomenological model.  
It consists of two low-energy and high-energy powerlaws with photon index $\Gamma_1$ and $\Gamma_2$, 
respectively, smoothly joined at some energy $E_{\rm b}$.
Among the physical models proposed for  a possible origin of the Band function it is worth mentioning the optically-thin 
synchrotron
model \citep[e.g.][]{tavani96}, in which the electron population of the relativistically expanding
shell is accelerated by internal shock collisions, eventually producing a supra-thermal powerlaw-like distribution 
in the tail of the 
Maxwellian.  This mixed electron population interacts with possible turbulent magnetic fields frozen in the plasma 
emitting synchrotron photons  and forming  synchrotron 
spectra \citep{rm05}.
The observed break in the energy distribution is thus naturally explained by the transition from the optically thick 
to optically thin emission regime.
However   the current  spectral analysis of both time-integrated \citep[e.g.,][]{crider97} 
and time-resolved  
\citep[e.g.,][]{crider98, frontera00, ghirlanda03} prompt GRB spectra has revealed some problems related 
to the synchrotron/SSC emission models.

Many theoretical efforts have also been performed in the thermal (photospheric) interpretation
\citep[e.g.,][]{thompson94, mr00, beloborodov10, toma11}.
From the observational point of view, \citet[][hereafer RP09]{rp09}, analyzed a sample of GRBs using archival
BATSE data and found that the time-resolved spectra can be fit by a high temperature ($10^9$K) blackbody (BB)  spectrum plus
a powerlaw (PL). Their model, albeit phenomenological, actually strongly points in favor of the presence of a
photospheric emission process at the origin of almost 50\% of the total emitted energy.
In fact, the low-energy threshold of BATSE prevented RP09 to use more detailed thermal models, but the good fits provided
by application of   the BB+PL  model could point in favor of the presence of a photospheric
(Compton-saturated) emission plus a second process giving rise to the PL-like hard X-ray emission.
As also argued by RP09, one of the advantages in considering thermal processes is to reduce the kinetic-to-radiation conversion efficiency, which is difficult to account by other theoretical models. However recently  
\cite{zy11} demonstrate that  the observed high efficiency is a drawback of the internal shock model, but is not an issue for the models that invoke dissipation of a Poynting flux. 

Another issue related to the GRB origin is the nature of the progenitor. 
Long GRBs are preferably observed at high redshift, with an average $z> 2$ \citep{jakobsson06, fiore07}.
They are concentrated in small, irregular galaxies and  show strong evidence of association  with 
Type Ic supernovae \citep{kkp08}. Both theoretical and observational
works point in favor of a  Wolf-Rayet star  with mass higher than $\sim$40 $M_{\odot}$ \citep{Raskin08} and 
stripped H-envelope as the most likely progenitor of long GRBs \cite[the collapsar model,][]{woosley11}.
These progenitor masses  are expected to leave, in their explosion,
a central remnant and  it is generally supposed that the GRB engine is powered by gravitational energy 
release of a torus of matter debris accreting onto a black hole. 
In some sense, a temporary \emph{microquasar} forms in the star after core-collapse, and consequently torus accretion
may power an expanding (relativistic) jet.
However, unlike the case of AGN or microquasar jets, the environment surrounding the outflow is not the interstellar
medium but the star photosphere which may play a significant role in dragging the jet kinetic energy.
The confinement level of the  expanding jet depends on the external environment; if the condition 
$\theta_{\rm J} M_{\rm J}  \la 2$ (where $\theta_{\rm J}$ and $M_{\rm J}$  are the jet opening angle and 
Mach number, respectively) is satisfied, the pressure equilibrium with the surroundings and
jet-structure is almost maintained, while in the opposite case strong shocks inside the jet may form and 
the overpressure allows a free expansion \citep{gehrels09}. Baryon-loading may also be a key ingredient 
in the determination of the jet bulk velocity.

In this context, it is worth mentioning
the theoretical work by  \cite{lazzati09}, which  used a numerically relativistic code \citep{fryxell00} 
to evaluate the evolution of a jet as it leaves a massive progenitor star after core-collapse and propagates to the photospheric 
radius, where radiation is eventually released. In their simulations, \cite{lazzati09} considered a 16 $M_{\odot}$  
Wolf-Rayet progenitor star evolved to pre-explosion 
and a jet with initial opening angle $\theta_0=10^{\circ}$ and Lorentz factor $\gamma_0=5$ at a distance
$R=10^9$ cm from the center of the star.
Although limited to this particular configuration,  the authors showed that the theoretical  
light curves are in good agreement with the observed ones. In addition, the  photospheric temperatures are in the
range 100-300 keV, in turn leading to values of the rest-frame  peak energy $\Ep$ consistent with observations.
However, also progenitors associated with much more massive stars (Population  III stars) cannot be excluded, 
at least for farthest GRBs. Indeed, supernovae associated with these stars with Helium core $M_{\rm c} > 100 M_{\odot}$, have been observed also in the closer Universe \citep{gy09}. Such progenitors, with initial mass in the 
range 130-260 $M_{\odot}$, have been shown to explode due to pair-instability (PI) \citep[e.g.,][]{Langer07,Woosley07,woosley11} 
with no compact remnant.
Thus GRBs could be the result of  helium burning in 
degenerated conditions, in which the burning occurs in deflagration regime [see details in \cite{cct10}].  Also in this case, \citep[e.g.,][]{imshennik99,b06} have shown that a shock runs away from the burning (reaction)
zone which leads to the development of an outflow (jet-like) structure. In terms of this model, 
the GRB spectrum would be originated from Compton upscattering of soft photons (probably  $\lax1$ keV)
in the hot sub-relativistic bulk outflow  region. 
The hydrodynamical simulations of \cite{cct10} show that the electron temperature of the hot corona region
can be as high as $T_{\rm e} \sim 10^9 K$.

The rapid temporal variability on time scale of order $\Delta T \la 1$ s and shorter,
observed during the GRB prompt phase, implies that the sources are compact  with 
size $R<c \Delta T\sim 10^{11} $ cm.  To avoid the problem of high compactness, that would imply high 
optical depth for pair production $\tau_{\gamma\gamma}\gg1$, a high Lorentz 
factor of the relativistic motion of the emitting plasma is the standard scenario \citep{piran99}.
Alternatively, to decrease $\tau_{\gamma\gamma}$,  it is needed to consider a much more extended volume over
which the photon field is distributed. 


In this Paper we offer a model to explain the GRB prompt spectral
formation in the context of {a photospheric scenario in which the main process
is the Comptonization of the relatively soft photons
of the star by a hot subrelativistic outflow  within  an area close to the photospheric 
radius (optical depth of  3-5) likely symmetric with respect to the rotational axis 
of the star.} 

We start from the theoretical and numerical results 
of the Comptonization problem reported in \citet[hereafter TMK97]{tmk97} and \citet[hereafter F08]{f08}, 
but for the case of an early subrelativistic bulk outflow phase  produced during the supernova explosion. 
This physical scenario is natural in the case of PI-SNe and it is foreseen in the case of an electromagnetic outflow \citep{lb03},
while in the collapsar model, the jet formed inside the star is assumed to be initially  relativistic. 
Whether this condition can hold, may depend on several parameters, such as
the initial jet energy, the chemical composition of the star envelope (in particular the presence or 
not of a H-envelope), the core angular momentum and the influence of magnetic torques \citep[]{gehrels09, woosley11}.

In \S \ref{Upscattering} we present the radiative transfer mathematical formalism,
necessary to investigate both the spectral formation in the subrelativistic phase, and the formation of the high-energy PL component.
In \S \ref{Amati_relation} we demonstrate that our upscattering model of  GRB radiation reproduces the Amati relation between the $EF(E)$ peak energy $E_p$ and the GRB radiated energy $E_{iso}$ or luminosity $L_{grb}$.  In \S \ref{temporal_var}  we put arguments for the observed high and slow temporal variabilities of GRB emission. 
In \S \ref{discussion}  
 we estimate  the resulting luminosity of GRB,   
 discuss a formation of  the high-energy  PL component and our scenario in the light of the {\it Fermi} results.
In \S \ref{conclusions} we draw our conclusions.

\section{Upscattering of soft photons  of the star in the hot plasma (Compton) cloud}
\label{Upscattering}

\subsection{Spectral formation in sub-relativistic outflow}
\label{bulk_compton}

The emergent spectrum produced by Comptonization of soft photons off a thermal electron population 
in the presence of bulk motion has been first formulated  by  \citet[hereafter BP81]{bp81a}
and later investigated by several authors \cite[e.g.,][]{cl82, colpi88}.
Depending on the sign of the divergence of the velocity field, the resulting spectrum can be harder
($\nabla \cdot {\bf V} < 0$) or softer ($\nabla \cdot {\bf V} > 0$) with respect to the case
of static medium \citep[][hereafter ST80]{st80}.
Examples of the first case (converging flow) are reported in TMK97 and F08,
while the second case (diverging flow) was investigated by \citet[][hereafter LT07]{lt99,lt07}.

The basic idea of the photon scattering effect in the case of bulk inflow and  outflow is presented in Fig. 1 of LT07: 
a photon emitted outwards near inner boundary and then scattered at a certain point by an electron moving 
with velocity ${\bf V}_1$, is
received by an electron having velocity ${\bf V}_2$ with
frequency $\nu _2 = \nu _1\left[1+\left({\bf V}_1-{\bf V}_2\right) \cdot{\bf n}/c\right]$ where ${\bf n}$ is a 
unit vector along the path of the photon at the scattering point.  In a diverging flow
$\left({\bf V}_1-{\bf V}_2\right)\cdot{\bf n}/c <0$ and photons are
successively redshifted, until they are scattered to an observer at infinity.
On the other hand, in a converging flow $\left({\bf V}_1-{\bf V}_2\right)\cdot{\bf n}/c >0$ and photons are blueshifted. This pure geometrical (Doppler) effect can be analytically described when the electron temperature
is very low, 
as the contribution of the thermal velocity vector ${\bf V_{\rm th}}$
is negligible compared with the dynamical (bulk) velocity vector ${\bf V_{\rm b}}$.
If, on the other hand, the plasma temperature is high enough so that  ${\bf V_{\rm b}}$ and ${\bf V_{\rm th}}$ 
become comparable, then  the emergent Comptonization spectrum is marginally modified by the presence
of the bulk motion.
A detailed analytical treatment of photon diffusion (Fokker-Planck approximation) in a converging flow
is reported in TMK97 and F08. 
Below we will start from the same equations treated by these authors, but we will consider the case of a fluid
which is subrelativistically expanding (diverging flow), namely we consider a velocity field with
$\nabla \cdot {\bf \Vb} > 0$.

\subsection{Formulation of the bulk motion problem. The main equation and its solution}
\label{sect_bulk_solution}

In this section, we provide the mathematical formulation of the problem required to compute the emerging spectrum due to Comptonization of soft photons by  electrons in a hot corona  bounced to the top of the star and moving outwardly with subrelativistic  velocity (see Fig. \ref{geometry}).
The spectrum can be derived using the Fokker-Planck expansion of the radiative transfer equation in the 
presence of  bulk motion (BP81, see also Eq. 13 in TMK97):

\begin{eqnarray}
\label{fp_equation}
\frac{\partial n}{\partial t}+{\bf \Vb}\cdot {\bf \nabla} n ={\bf \nabla} \cdot \left[\frac{c}{3k}{\bf \nabla} n\right] +
\frac{1}{3}\nu \frac{\partial n}{\partial \nu} {\bf \nabla} \cdot {\bf \Vb} +
\\\nonumber
\frac{1}{\nu^2} \frac{\partial}{\partial \nu} \left[\frac{kh}{\me c} \nu^4\left(
n + \frac{\kte+ \me \Vb^2/3}{h} \frac{\partial n}{\partial \nu}\right)\right]+ j(r,\nu),
\label{main_fp_equation}
\end{eqnarray}
where ${\bf \Vb}$ is the bulk velocity field of the outflow,  $\kte$ its electron temperature, $k(r)=\Ne (r)\sigma_T$ 
is the inverse of the scattering mean free path, $n \equiv n(r,\nu)$ is the zero-moment occupation number 
of the specific intensity and  $j(r,\nu)$ is the source term.

 Let $N(r)=N_0(r_0/r)^{\mu}$ be the radial number density profile of the bulk outflow, with radial
outward velocity given by
\begin{equation}
\beta(r)=\beta_{0}\left(\frac{r_0}{r}\right)^{2-\mu},
 \label{velocity_prof}
 \end{equation}
which is easily derived from the continuity equation in  spherical geometry. 
 Here $\beta_0=V_0/c$ is the dimensionless outflow  velocity  with respect to speed of light at the bottom of the shell. 
 Thomson optical depth of the flow from some radius $r$ to infinity is given by

 \begin{equation}
\tau=\int^\infty_{r} \Ne \sigma_T dr' = \tau_0(r_0/r)^{\mu-1},
\label{tau}
\end{equation}
where $\tau_0$ is the optical depth at the bottom of the shell.
Using the given radial density profile together with definitions (\ref{velocity_prof}) and (\ref{tau}), the steady-state 
Fokker-Plank (diffusion) equation  (\ref{fp_equation}) becomes 
\begin{equation}
\frac{1}{\Theta} L_{\tau}n +\delta\cdot x \frac{\partial n}{\partial x}+ \frac{1}{x^2} 
\frac{\partial}{\partial x}\left[x^4\left(\fb \frac{\partial n}{\partial x} +n\right)\right]=- \frac{j(\tau, x)}{\Theta k},
\label{red_kin_eq}
\end{equation}
where  

\begin{equation}
\delta=  \frac{\mu}{3 \Theta(\mu-1)}\frac{\beta_0}{\tau_0}\left(\frac{\tau}{\tau_0}\right)^{(3-2\mu)/(\mu-1)}.
\label{bulk_parameter}
\end{equation}

In equation  (\ref{red_kin_eq}), $L_{\tau}$ is the space operator \citep[see][]{tkb03}, 
$x\equiv h\nu/(\me c^2)$ and  $\Theta \equiv  \kte/(\me c^2)$ are dimensionless photon energy and electron temperature,  respectively, while $\fb=1+(\Vb/c)^2/(3\Theta)$.

We note that $\delta$ in equation  (\ref{bulk_parameter}) is constant only for $\mu$=3/2, which corresponds to the free-fall 
velocity profile case (BP81), while for $\mu \neq 3/2$ it is a function of  the optical depth $\tau$. 
For the particular case of constant outflow velocity ($\mu=2$), equation (\ref{red_kin_eq}) becomes
\begin{equation}
\frac{1}{\Theta} \left[\frac{1}{3}  \frac{\partial^2 n}{\partial \tau^2}-(\frac{4}{3\tau}-\beta_0)\frac{\partial n}{\partial \tau}\right]
+\frac{2\beta_0}{3\tau}x\frac{\partial n}{\partial x}+
\frac{\partial}{\partial x}\left[x^4\left(\fb \frac{\partial n}{\partial x} +n\right)\right]=-\frac{j(\tau, x)}{\Theta k}.
\label{rte_velconst}
\end{equation}

The dependence of $\delta$ on  $\tau$ does  not allow  to find a solution of the  equation 
(\ref{rte_velconst})  with the method of variable separation  (space and energy).
 However, LT07 demonstrate that for a constant velocity profile, $\delta$ can be replaced by 
 its value at some effective optical depth $\taueff$, which is a fraction of total optical 
depth $\tau_0$ of the expanding shell.
Thus, replacing $\tau$ with $\tau_{eff}$ in equation (\ref{bulk_parameter})
we can now rewrite equation  (\ref{red_kin_eq}) in the form
\begin{equation}
({\cal L}_{\tau} + {L}_x) n(x,\tau)=-\frac{\varphi(x)}{x^3}s(\tau),
\label{eigen_problem}
\end{equation}
where ${L}_x$ and  ${\cal L}_{\tau}=\Theta^{-1}L_{\tau}$   are the energy and  space  differential operators (see these  operators in Eq.   \ref{red_kin_eq}  for $\delta=const$ and Eq.  \ref{rte_velconst}  respectively).
  $\varphi(x)$ is the occupation number 
of the seed photons, and $s(\tau)$ their spatial 
distribution  in the bounded medium.
The solution of equation (\ref{eigen_problem}) can be conveniently expressed as a  series
 $n(x,\tau)=\sum c_k N_k(x) R_k(\tau)$, where $R_k(\tau)$ are the eigenfunctions of the space operator 
  $L_{\tau}$ and $c_k$ are the expansion coefficients over the seed 
photon spatial distribution $s(\tau)$.
We calculate the  Comptonization spectrum using  only the first
term $k$=1 of the series (TMK97), which is obtained solving   the equation
\begin{eqnarray}
\label{energy_op_full} 
\fb x^2\frac{d^2N(x)}{d^2x}+(x^2+4\fb x+\delta x)\frac{dN(x)}{dx} \\ \nonumber
+(4x-\gamma)N(x)=-\gamma \frac{\varphi(x)}{x^3},
\end{eqnarray}
where
\begin{equation}
\delta=\frac{2 \beta_0}{3 \taueff \Theta},
\label{delta_def}
\end{equation}
is given for $\mu=2$,  $\tau_{eff}=<\tau>$ (see Eq. \ref{bulk_parameter}) and  $\gamma=\lambda_1^2/\Theta$. Here $\lambda_1^2$ is the first  eigenvalue of the 
space operator, $L_\tau$ specifically
\begin{equation}
L_\tau R(\tau)+\lambda_1^2 R(\tau)=0.
\label{eigen_value_problem}
\end{equation}

For the particular case $(\Vb/c)^2\ll 3\Theta$, $\fb\sim1$ and it is possible to express the analytical
solution of equation  (\ref{energy_op_full}) in terms of the convolution of the seed photon spectrum
$\varphi(x)$ with the Green function (GF) of the energy operator $L_x$ according to

\begin{equation}
F_{\rm tb}(x)=\int^\infty_0 G(x,x_0) \varphi(x_0) dx_0,
\label{convolution}
\end{equation}
where the GF is defined as 
\begin{eqnarray}
G(x,x_0)=\frac{C_{\rm N} e^{-x}}{x_0 \Gamma(q)}
\cases{\displaystyle 
{\left (\frac{x}{x_0}\right)^{\alpha+3-\delta}  {_1F_1}(\alpha, q, x) I(x_0, \alpha,\delta)},
{\rm for}~x \leq x_0;
\cr
{\displaystyle\left(\frac{x}{x_0}\right)^{-\alpha} { _1F_1}(\alpha, q, x_0) I(x, \alpha,\delta), }
~~{\rm for}~~x  \geq x_0,}
\label{green_func}
\end{eqnarray}
where
\begin{equation}
 I(x, \alpha,\delta)= \int_0^{\infty}e^{-t} (x+t)^{\alpha+3-\delta} t^{\alpha-1}dt, 
\end{equation}
and 

\begin{equation}
q= 2\alpha+4-\delta,
\end{equation}
while ${ _1F_1}(a, b, z)$ is the Kummer confluent hypergeometric function.
Note the change of sign of $\delta$ with respect to the same GF reported in F08 because
here we are considering a diverging flow.
The energy spectral index $\alpha$ of the GF is given by  

\begin{equation}
\alpha=-\frac{3-\delta}{2} + \left[\left(\frac{\delta+3}{2}\right)^2+\gamma\right]^{1/2},
\label{alpha_tb}
\end{equation}
and the normalization $C_{\rm N} =\alpha(\alpha-\delta+3)$  is chosen in order to conserve the total
photon number, namely
 \begin{equation}
\int_0^\infty {G(x,x_0)}\frac{dx}{x} = {{1} \over {x_0}}.
\label{green_norm}
\end{equation}

It is worth pointing out that $\lambda^2_1 \propto 1/\tau^2_0$, where $\tau_0$ is the total Thomson optical depth
of the shell, so that  $\gamma \ll 1$ corresponds to $\Theta \tau^2_0 \gg 1$, namely  for the case of  saturated Comptonization. 
In the case of a pure thermal motion ($\delta$ = 0) this condition  implies $\alpha \sim 0$ (see Eq. [\ref{alpha_tb}]), and the emergent spectrum obtained from equation  (\ref{convolution}) has a Wien-like shape, no matter which is the spectral distribution $\varphi(x)$ of the seed photons. On the other hand, if bulk motion is present, then $\alpha \approx \delta$.

In Figure  \ref{plot_green_func} we present a typical example of the shape of the GF for different values of 
the bulk parameter $\delta$ defined in equation (\ref{delta_def}), for the case of soft photon monochromatic injection 
($x_0 = 0.1$) and $\gamma \ll 1$.  While 
in Figure \ref{conv_spectrum} we show a convolution of the Green function with a BB seed spectrum (see 
Eq. \ref{convolution}).
It is evident  that the important role is  played by bulk Comptonization in diverging flow in determining the
slope of the high-energy wing and  the position  of the high-energy cut-off, both quantities modify  the 
position of  the peak in the EF(E) spectrum  
(see Fig. 1 in F08 for comparison with the case of converging flow).

As already outlined above, these analytical results can be obtained for the case $\fb$=1 (see Eq.  \ref{energy_op_full}). On the other hand,  if $\fb > 1$, the equation has to be solved numerically,
and we report the details of the numerical solution in Appendix \ref{sweep}.

\subsection{High energy power-law component in the GRB prompt emission}
\label{sect_powerlaw}

The solution of equation (\ref{energy_op_full}) clearly shows that above the
energy peak $\Ep$ in the EF(E) diagram,  the spectrum has an exponential rollover 
(see Fig. \ref{conv_spectrum}). This spectral feature is  characteristic of Comptonization 
both in a pure thermal or subrelativistic moving cases.
For  a thermal Comptonization, the cut-off energy $\Ec$ is dictated only by the plasma 
electron temperature, with $\Ec \gax 2 \kte$, while when bulk motion is present (inflow or outflow), 
 the bulk-parameter $\delta$ also concurs in dictating  the $\Ec$-value (see Figs. \ref{plot_green_func}
and \ref{conv_spectrum}). In both cases, the high-energy exponential rollover
arises from the presence of the thermal recoil-effect $\propto N(x)$ in the radiative transfer equation
modified by the term $\propto \delta$ (see Eq. \ref{red_kin_eq}).

However, it is a well established observational result that most (albeit not all) of the observed prompt GRB spectra 
 show an extended PL above the peak energy in the $EF(E)-$diagram [where $F(E)$ is the energy spectrum], which is phenomenologically described by
the second index $\Gamma_2$ of the Band function. This high-energy feature cannot be reproduced by the solution of  Equation  
(\ref{energy_op_full}), and a second component to describe the broad-band spectrum is required.
One possibility is to simply  add a PL to the thermal spectral component, as was done, e.g., by
RP09, who fitted the BATSE spectra with a BB+PL model.
From a physical point of view, this model can imply a different 
origin for the apparently thermal
and non-thermal part of the GRB spectrum. It is also well known that the inclusion of a simple PL in broad-band 
spectral model can introduce some biases, because this component, intended to be used for describing
only part of the spectrum (usually at high energies), actually covers the whole spectral range, producing
undesired effects such as  overestimation of the flux at low energies
(the effect getting stronger for steeper photon index $\Gamma$) \citep[Frontera et al. 2012, in preparation;][]{Ghirlanda07}.

Here, we investigate a different possibility;  there  is a general theorem that the solution of an equation with linear  differential operator and a source term (such as Eq. \ref{energy_op_full}) is obtained by the convolution of the 
GF of the given operator with the  source function itself, according to Eq. (\ref{convolution}).
Examples for this convolution, in addition to the results of previous section,  are reported for the case of pure thermal 
Comptonization (ST80), thermal plus bulk Comptonization (TMK97, F08), and Comptonization in strong-magnetic field \citep{bw07}.
The resulting energy  spectrum can be written in the general form as 
\begin{equation}
 F(E)=\frac{1}{A+1} [\varphi(E) + A \times \varphi(E_0) \ast G(E,E_0)],
\label{bmc_equation}
\end{equation}
where the two terms on the right-hand side represent the direct and Comptonized (convolved)
part of the seed spectrum, respectively. 
The weighting factor $A$ is related to the physical and geometrical configuration
of the system. The high-energy behavior of the spectrum depends on the kernel
$ G(E,E_0)$; for example, if it is a broken PL with no cut-off, then the
convolution with $\varphi(E)$ (see Eq. \ref{convolution}) produces  a PL (BMC model
in the XSPEC package). 

Thus the observed GRB prompt spectra, phenomenologically
described by the Band function, can be obtained by an equation of the form (\ref{bmc_equation})
with $A \gg 1$, the source term $\varphi(E)$ obtained by the solution of equation (\ref{energy_op_full}) and using a broken-PL GF with no recoil-term.
The latter can be conveniently described by the following form:
\begin{equation}
G_{bpl}(x,x_0)=\frac{\alpha_{\rm b}(\alpha_{\rm b}
+3)}{x_0 (2\alpha_{\rm b}+3)}
\large
\left\{ \begin{array}{ll}
\left(\frac{x}{x_0 }\right)^{\alpha_{\rm b}+3} & \mbox{ for $x \leq x_0$;}\\ 
\left(\frac{x}{x_0}\right)^{-\alpha_{\rm b}}  & \mbox{for $x \geq x_0$},
\end{array}
\right.
\label{green_func_up}
\end{equation}
where the subscript $b$ has been used to avoid confusion with the index $\alpha$ of
the GF related to the thermal plus bulk energy operator (Eq. \ref{green_func}).
Note that this is the GF of the energy operator in equation  (\ref{red_kin_eq}) with
$\fb$=1, $\delta$=0 and the recoil term $\propto x^4~N$ in square brackets dropped \citep{st80},
which leads to the equation 
\begin{equation}
x^2\frac{d^2W}{dx^2}+4x\frac{dW}{dx}- \alpha_{\rm b}(\alpha_{\rm b}+3) W(x)=- \alpha_{\rm b}(\alpha_{\rm b}+3) \varphi(x),
\label{up_scattering}
\end{equation}
where $\varphi(x)=N(x)$ is obtained by solving equation (\ref{energy_op_full}).
However, the spectrum obtained as a solution of equation  (\ref{energy_op_full}) has no analytical representation, 
so performing a numerical convolution according to equation  (\ref{convolution}) where $G_{bpl}(x,x_0)$ is
given by equation  (\ref{green_func_up}) and $\varphi(x)$ by $N(x)$ in equation  (\ref{energy_op_full})
can be time-consuming in terms of CPU computation.
We thus again numerically solved equation (\ref{up_scattering}) using a finite-difference
method. From the mathematical point of view, there is perfect equivalence between the analytical solutions  given by the  convolution above described and  direct  
numerical solution of the equation. In our model $\alpha_{b}$ is a free parameter, related to the high-energy PL photon index $\Gamma_2$
of the Band function,  by $\alpha_{b}=-\Gamma_2 -1$.

To better explain this behavior, in  Fig. \ref{plot_bmc} we show two spectra 
 obtained from the \bmc\ model of the \xspec\ package (TMK97), where the source function
$\varphi(E)$ is assumed to be a BB occupation number. Note that when $A \gg1$, only the second (convolution) term contributes to the spectral formation, and the low-energy and high-energy parts
of the spectrum smoothly join at the peak energy, in a \emph{qualitative} way which
is very similar to the Band function.

This is actually the approach we adopted to develop our model: we first solved the Comptonization
equation in the presence of subrelativistic bulk motion (see Eq. \ref{energy_op_full}) and then
we convolved it with the broken-PL GF of equation (\ref{green_func_up}), setting $A \gg 1$ in  equation (\ref{bmc_equation}).
In  Appendix \ref{xspec_model} we describe the numerical methods used to address these issues.

\section{The Amati's relation and its interpretation}
\label{Amati_relation}

{ In spite of the still open discussion about the impact of possible selection effects on the correlation between
the peak energy   of 
$EF(E)$ diagram $E_p$
and
the GRB radiated energy or luminosity 
\citep[Amati relation, AR,][]{amati02},  it is a matter of fact that all GRBs with 
known redshifts, but one (GRB 980425), follow this relation 
[see, e.g., \cite{amati09} and references therein]. In addition, it is found that also within single GRBs, this relation holds 
\citep[Frontera et al. submitted;][]{Ghirlanda10}.

Thus it is crucial to understand the origin of this relation},
namely which is the mechanism (at first glance universal)
which gives rise to the relation $\Ep \propto E_{\rm iso}^{1/2}$ where $\Ep$ and $E_{\rm iso}$
are the peak energy   of 
$EF(E)$ diagram and  isotropic radiated energy (fluence)
of GRB  respectively during the prompt phase.

The main issue to be investigated is to check how the parameters of our model concur
in determining the energy peak value $\Ep$ in the $EF(E)$ diagram and the total luminosity $L_{\rm grb}$ which integral over prompt emission time is $E_{\rm iso}$.

In  Fig. \ref{eeuf_grbcomp_delta0.5_alpha1.5-30}  we show the resulting spectra obtained by
fixing the parameters of the early sub-relativistic phase and changing the spectral index 
$\alpha_{\rm b}$ of the GF defined by equation (\ref{green_func_up}), which is responsible for the high-energy
PL component. 
Given that $\Ep$ very slightly depends (within a factor less than two) on different $\alpha_{\rm b}$-values,
we  fix $\alpha_{\rm b}$ about 1.5.  In Fig. \ref{spec_parameters} we show the theoretical EF(E) spectra obtained 
using our model alternatively changing $\beta_0$, $\kte$, $\delta$ and $\ktbb$, namely the parameters which 
characterize the subrelativistic phase. 
 In addition we also show an example of  the spectrum obtained using the solution of Eqs.  (\ref{energy_op_full}) and (\ref{up_scattering}) compared  them with the Band  function in  Fig. \ref{efe_grbcomp_band}.
{\it The result is that the leading quantity in influence of  $\Ep$ is the electron temperature
$\kte$.}
This result is not surprising because it is evident that in any photospheric model with spectral emission dominated
by Comptonization, the electron temperature represents a key ingredient, both in  a pure thermal and dynamical cases (for $\beta\ll 1$).
On the other hand, it is worth pointing out that $\Ep$ is independent of  
values of the BB-like seed photon temperature $\ktbb$.
Thus  different values of $\ktbb$ do not change $\Ep$, but they  determine the total (resulting) luminosity, as we discuss below.

The question that naturally  arises is whether the observed dependence of $\Ep$  on $E_{\rm iso}$ (or on luminosity $L_{\rm grb}$)   
and its intrinsic dispersion 
 \citep{amati08} are the result of a multi-parametric combination of the subrelativistic phase plus
the effect of the non-thermal process or 
is the fundamental effect of $\gamma-$  ray emission.

{ As an example, in Figure~\ref{grb990705} we report the results of the fit using our model (see Appendix B) of the time averaged prompt 
emission spectrum of GRB\,990705 obtained with {\it Beppo}SAX and, for comparison, with the Band  function.}
With our model, the best-fit parameters are $\ktbb= 1.8^{+  1.3}_{-  1.4}$ keV, $\kte=94^{+ 30}_{- 10}$ keV,
$\delta= 0.32^{+  0.09}_{-  0.03}$, $\alpha_{b}=1.50^{+  0.40}_{-  0.17}$ and photospheric radius $\rph= 1.5 \times 10^{13}$ cm, with 
\chiq/dof=168/182.
We fixed $\gamma=5\times 10^{-3}$ (as a $\gamma$-value strictly equals to zero produces computer
underflow) and outflow velocity $\beta_0=0.2$.
The Band function on the other hand provides a low energy  photon index $\Gamma_1=-0.86^{+  3.91}_{- 1.24}$ and 
$E_0 =292^{+18}_{-22}$ keV with \chiq/dof=167/182. In this case however the high-energy index is not  constrained and  we fixed $\Gamma_2=-2.5$ (note that because of sign convention $\alpha_{\rm b}= - \Gamma_2 -1$).

The small values of the best fit parameters  of our model $\gamma$ and  $\delta$  correspond to the case of the saturated Comptonization for which the resulting index $\alpha\ll1$ (see Eq. \ref{alpha_tb}). 
\cite{st80}, \cite{st85} [and see also \cite{ct95}] derive the formula for the Comptonization enhancement  factor $\eta_{\rm comp}$ which is  a ratio of the resulting luminosity, that is in our case, the GRB luminosity  $L_{\rm grb}$ 
to the injected luminosity of soft photons $L_{\rm soft}$. 
Namely
\begin{equation}
\eta_{\rm comp}=\frac{L_{\rm grb}}{L_{\rm soft}}=q_{x_0}(\alpha)x_0^{\alpha-1},
\label{enhancement_factor}
\end{equation}
where 

\begin{equation}
q_{x_0}(\alpha)=\frac{\alpha(\alpha+3)\Gamma(\alpha+4)\Gamma(\alpha)\Gamma(1-\alpha)}{\Gamma(2\alpha+4)} (1-x_0^{1-\alpha}),
\label{q_alpha}
\end{equation}
and $x_0=2.7\ktbb/\kte$.
Thus when $\alpha\ll1$
\begin{equation}
\eta_{\rm comp}\propto \kte.
\label{enhancement_factor_m}
\end{equation}

But the flux of soft photons illuminating the hot spot (Compton cloud)  $L_{\rm soft}$ is 
\begin{equation}
L_{\rm soft}= \pi B_{\rm soft}S_{\rm hs}
\label{L_soft}
\end{equation}
where $B_{\rm soft}$ is the intensity of the blackbody radiation of the star and  $S_{\rm hs}$ is the surface area of Compton hot spot.

The thermal wave  propagates in the hot spot  with plasma velocity $V_p$ whose value can change from one GRB to another. 

As a consequence for each GRB $S_{\rm hs}\propto (V_pt)^2$ and then  
\begin{equation}
S_{\rm cc}\propto V_p^2\propto kT_{\rm p}= \kte.
\label{hot_spot_area}
\end{equation}

Thus  the luminosity of the GRB hot spot $L_{\rm grb}$ should be
\begin{equation}
L_{\rm grb}=\eta_{\rm comp}L_{\rm soft} \propto ( \kte)^2.
\label{L_subrel}
\end{equation}

In order to calculate the GRB fluence $E_{\rm iso, grb}$ one should integrate $L_{\rm grb}$ over the GRB prompt emission time  $T_{pr}$, namely  
\begin{equation}
E_{\rm iso}=\int_0^{T_{pr}}  L_{\rm grb}(t)dt.
\label{fluencel}
\end{equation}
If the time-scale of the GRB prompt emission and its shape  is more less the same for any burst then  
$\Ep$ ($\propto \kte$) is $E_{\rm iso}^{1/2}$ which is precisely seen in the Amati  relation [\cite{amati02} and \cite{amati06}]. 
The observable scattering of points in the Amati relation along correlation 
$E_p\propto E_{\rm iso}^{1/2}$ can be caused by the spread of the parameter values that characterize the outflow evolution of each GRB like the bulk velocity $V_b$ and the spectral index of the relativistic phase $\alpha_b$  
(see Eq. \ref{green_func_up}  and Fig. ~\ref{efe_grbcomp_band}).
Another reason for the spread in the Amati relation is the different temperature of the seed (soft) photons,  
which is related to the emission of the star itself.


\section{Temporal variability}
\label{temporal_var}

One of the most established observational results of GRBs since the beginning is the high temporal variability 
of their light curves \citep{meegan92}, which can extend down to 10$^{-3}$ s. More recently, in addition
to the short time variability, also a slow time variability, with time scales from several seconds to 
$\sim 100$ s has been discovered, with the GRB prompt light curves being the superposition of both components
\citep[e.g.,][]{vetere06,Gao11}. In particular, \citet{vetere06}, by performing
the systematic analysis of a sample of GRBs detected with the WFCs aboard
\sax\ also showed that the slow component is more pronounced at lower energies (2-10 keV) than at high energies (10-26 keV). Accumulating evidence from the \swift\ era suggests that at least the slow component
of the GRB variability  is closely related to the central engine \citep{liang06}, which was confirmed also by other
investigations \citep{Gao11}.  

\begin{figure}[]
\includegraphics[width=16cm,height=14cm]{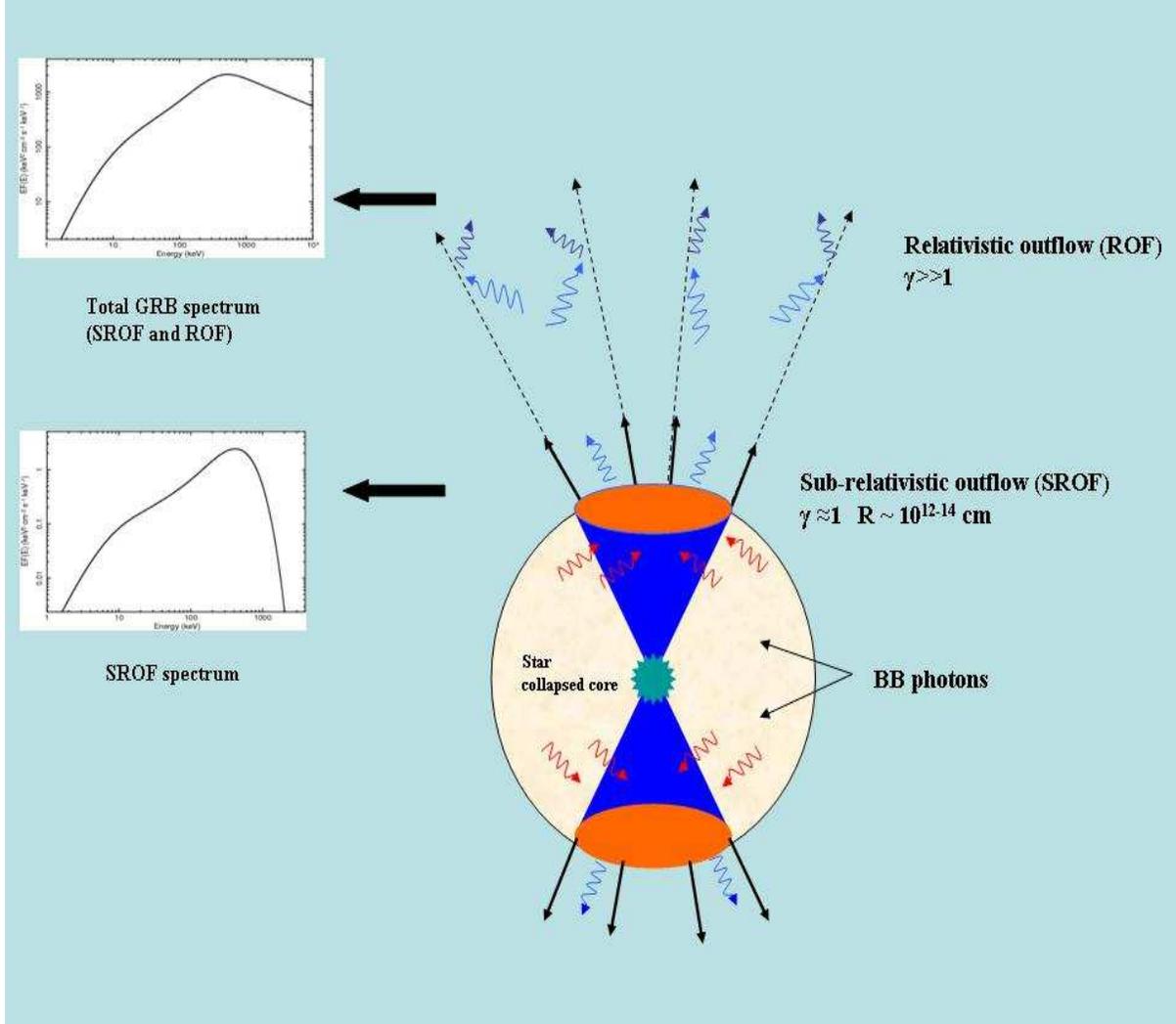}
\caption{Schematic view of the explosion of a massive star and associated emergent spectral components. 
At the first stage, the  spectrum is formed by Comptonization of blackbody-like seed photons ({\it red arrows}) placed at the bottom of
a hot sub-relativistically outflow (SROF or Compton cloud)  on the top of the star photosphere ({\it orange region}). 
The emergent Comptonization spectrum 
({\it bright-blue arrows}) is obtained using the solution of  Eq.  (\ref{energy_op_full}). 
The Compton cloud can be accelerated by the underlying radiation pressure of
the star photosphere leading to the relativistic outflow (ROF).
Inverse Compton scattering of the Comptonized  photons of the subrelativistic
phase  off  non-thermal relativistic electrons   of ROF  can be the origin of the extended power-law observed above the peak in the EF(E) diagram ({\it dark-blue arrows}).}
\label{geometry}
\end{figure}

\begin{figure}
\includegraphics[width=12cm,height=14cm,angle=-90]{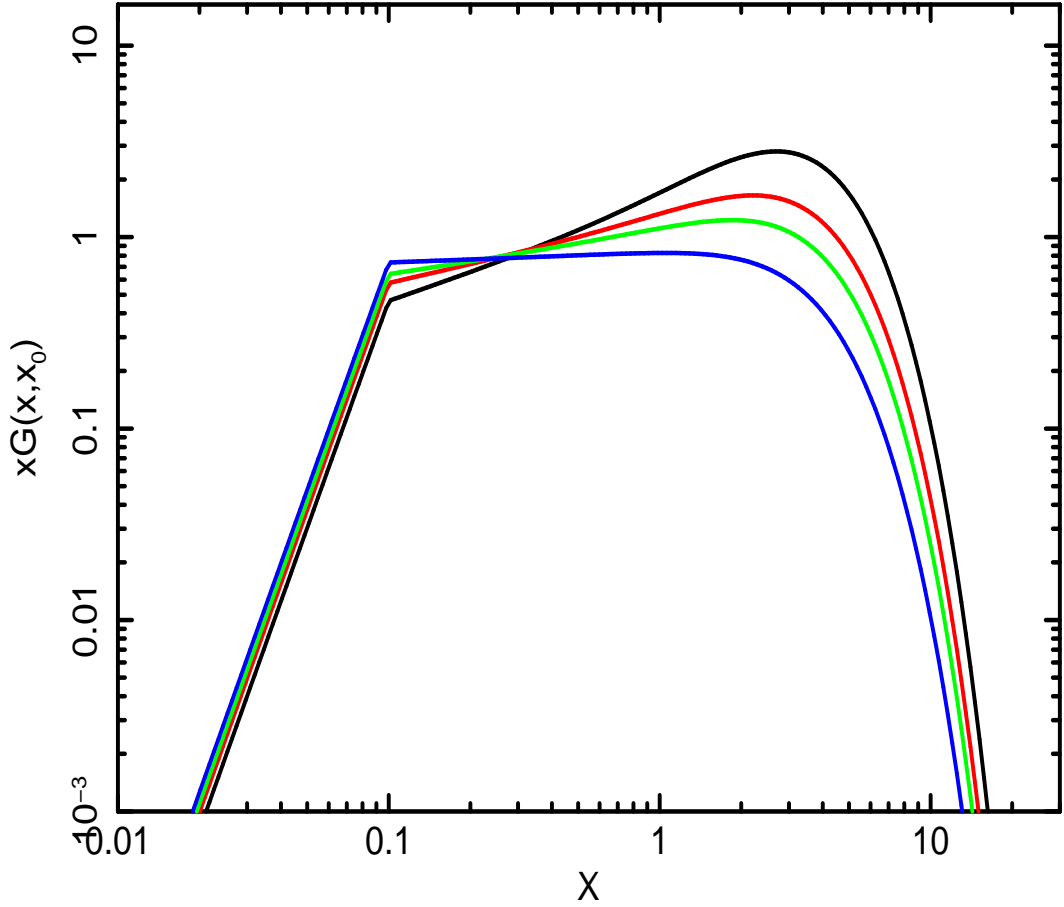}
\caption{Evolution of the Green function  (\ref{green_func}) for the specific case of saturated Comptonization 
($\gamma \sim 0$) and different values of the bulk parameter $\delta$. Black, red, green and blue correspond 
 to $\delta=$0.5, 0.7, 0.8 and 0.95 respectively. The horizontal scale is defined in units $x\equiv E/\kte$.}
\label{plot_green_func}
\end{figure}

\begin{figure}
\includegraphics[width=12cm,height=14cm,angle=-90]{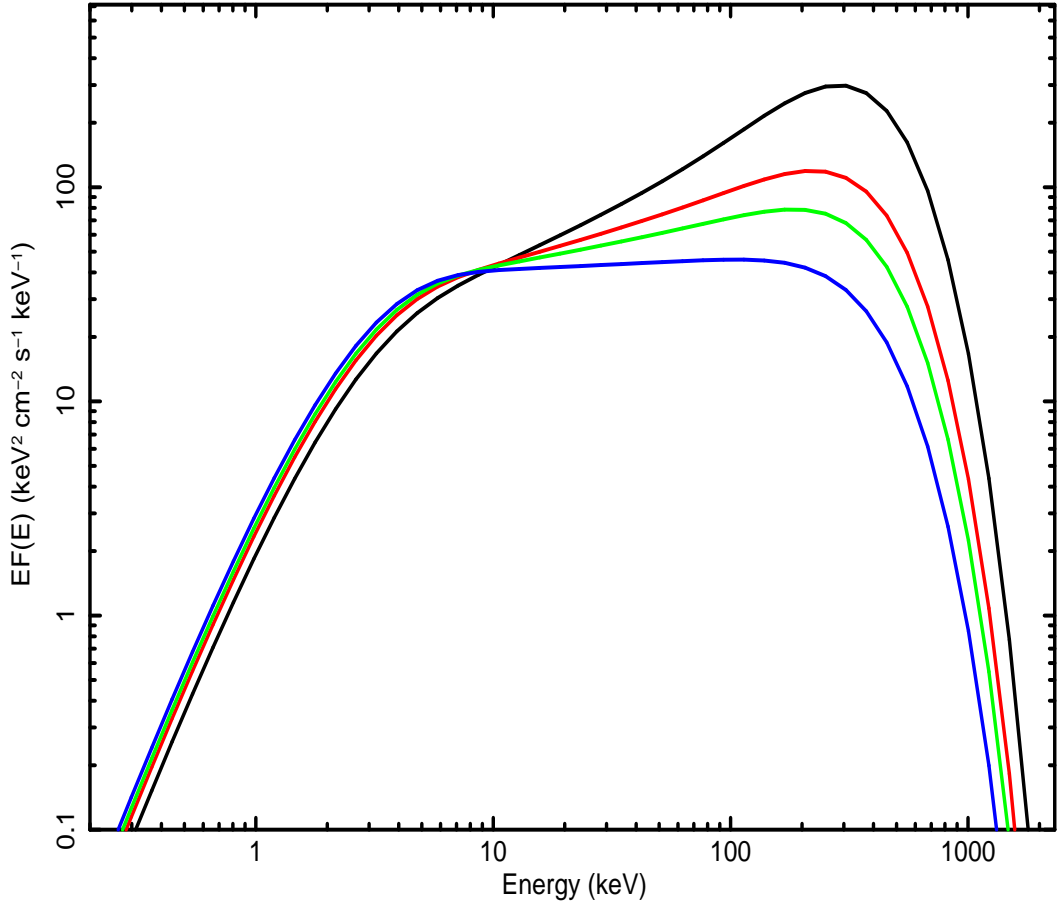}
\caption{Saturated Comptonization spectra ($\gamma \sim 0$) modified by the presence of subrelativitic bulk motion obtained from the  numerical convolution of the 
Green function (\ref{green_func}) with a blackbody spectrum of temperature $\ktbb=1$ keV. Different colors represent different values of
the bulk parameter $\delta$ according to Fig. \ref{plot_green_func}. A plasma temperature of $\kte=100$ keV is assumed.}
\label{conv_spectrum}
\end{figure}

\begin{figure}
\includegraphics[width=12cm,height=14cm,angle=-90]{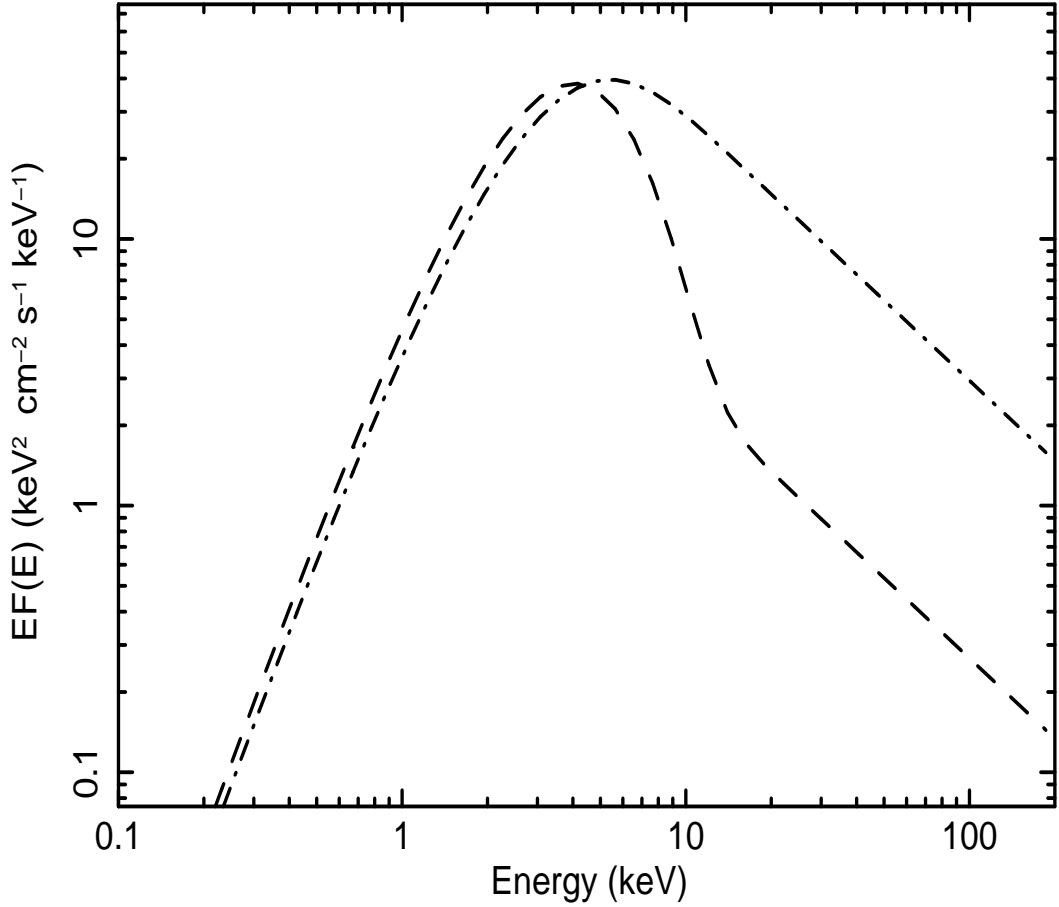}
\caption{Emergent spectra obtained as the sum of a BB spectrum plus its convolution  
with a broken power-law Green function (see Eq. \ref{green_func_up}) and with
weighting factors A $\gg$ 1 ({\it dotted-dashed line}) and A=0.1 ({\it dashed line}), according to equation (\ref{bmc_equation}).  In both cases the BB temperature is 1 keV and 
the Green function spectral index $\alpha_{\rm b}$=2.}
\label{plot_bmc}
\end{figure}

\begin{figure}
\includegraphics[width=12cm,height=14cm,angle=-90]{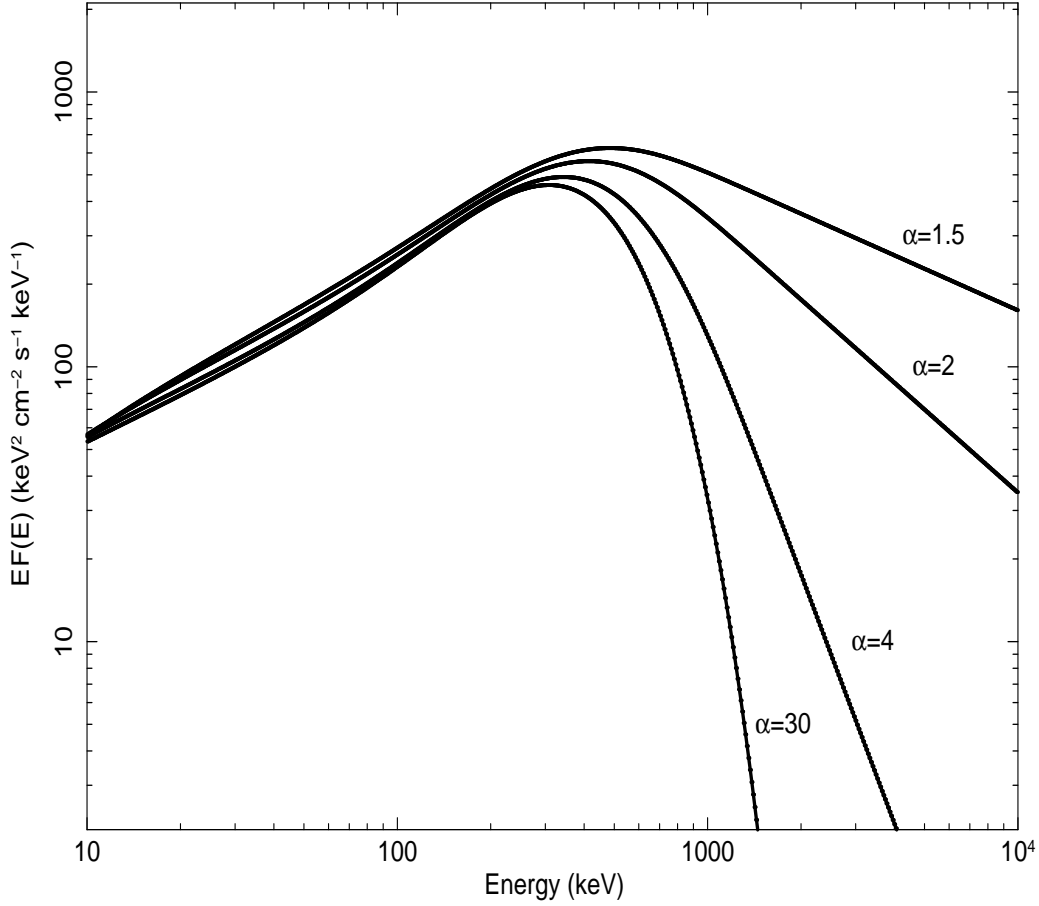}
\caption{Emergent spectra obtained by the solution of Equations (\ref{energy_op_full}) and (\ref{up_scattering})
for different values of the up-scattering Green function index $\alpha_{\rm b}$ determined  by Eq. (\ref{green_func_up}). Other parameters
of the model are $\ktbb=1$ keV, $\kte=100$ keV, $\delta=0.5$, $\beta_0=0.2$, and $\gamma \sim 0$.} 
\label{eeuf_grbcomp_delta0.5_alpha1.5-30}
\end{figure}

Now we demonstrate that, in our scenario, the slow component with temporal  variability of order 1-100 seconds is  related to the Compton cloud while  the fast variability component with scale of $\ll 1$ s is presumably related to the relativistic outflow (jet). 
In order to estimate the time scale of emergent radiation from the hot Compton cloud one should take a difference 
of the arrival time from the outskirt $t_{\rm out}$ and center of Compton cloud $t_{\rm c}$ (see Fig. \ref{geometry}). Namely
\begin{equation}
\Delta t=t_{\rm out} - t_{\rm c},
\label{time_interval}
\end{equation}
where 
$t_{\rm out}= \rho_{\rm out}/c$, 
$t_{\rm c}= d/c$ 
and  $d$, $\rho_{\rm out}$ are the distances from the observer to the  center and outskirt of the hot spot 
respectively (see Fig. \ref{geometry}).
Thus
\begin{equation} 
\rho_{\rm out}= \sqrt{R^2\sin^2\varphi +[d+(R-R\cos\varphi)]^2} 
\label{outskirt_distance}
\end{equation}
where $R$ is the radius of the star and $\varphi$ is angle between the vector directed from the center of the star to outskirt point and vector directed towards the observer. 
Then for $R\ll d$ we find that 
\begin{equation}
\Delta t=t_{\rm out} - t_{\rm c}\approx R(1-\cos\varphi)/c
\label{time_interval_mod1}
\end{equation}
and  for $\varphi\ll \pi/2$ we obtain
\begin{equation}
\Delta t\approx \varphi^2R/2c.
\label{time_interval_mod2}
\end{equation}

For example if $\varphi<0.1$ (or $\varphi<5^o$ in degrees) then the variability  time scale of the emergent radiation observed from the Compton cloud 
should be
\begin{equation}
\Delta t< 15(R/10^{14}{\rm cm})~ {\rm s}.
\label{time_interval_estimate}
\end{equation}
which is a typical long time scale of the GRB prompt emission (see above).

On the other hand short time scale variability ($t\ll 1$ s) is presumably related to relativistic outflow (see section 
\ref{sect_powerlaw} and below).   Variations of X-ray$-\gamma$ radiation in the outflow  in time scale $T$ 
detected by the Earth observer as 
\begin{equation}
T_0=T[1-(V_b/c)\cos\theta]
\label{time_scales_observer_vs_jet}
\end{equation}
where $V_b$ is the outflow velocity in the relativistic phase and   $\theta$ is angle between the direction of the outflow motion and the direction to the Earth observer [see \cite{rl79}]. If the observer sees the outflow motion at $\theta\ll1$ then the time scale is very short
\begin{equation}
T_0=T/(2\Gamma^2_L)
 \label{time_scales_observer_vs_jet_m}
\end{equation}
where $\Gamma_L$ is the Lorentz factor of the outflow(jet).  
It means that the short variability time scales should be only seen when the jet is directed towards the observer. 
The short temporal  variability seen in GRB of order of $10^{-3}-10^{-1}$ s can be considered a signature of the outflow for which time scales of X-ray$-\gamma$ variations  of order  of seconds are seen by the Earth observer as    
$10^{-3}-10^{-1}$-s variations. 
In other words the long time scales of order 10-100 s (intrinsic variabilities of X-ray radiation in the hot spot region) are related to the Comptonized radiation of the hot spot  in the subrelativistic phase  while the short time scales of order $10^{-3}-10^{-1}$ s, as  seen by the Earth observer,  are originated mainly in the relativistic outflow phase {\it similar to that in the standard fireball scenario} \citep[e.g.][]{m06}.

{\it Using the emergent spectra itself one cannot distinguish the contribution of Compton cloud and the outflow in the spectrum below the peak energy $\Ep$ because the Comptonization spectrum is wider than the upscattering Green function of the outflow},  at least for $\alpha_b\gax 1.5$ (see section \ref{sect_powerlaw}).
The short and long time variabilities  should be seen in the same energy range at $E\lax \Ep$ but the short one should be mostly  observed alone  at high energies (above  $\Ep$) where the pure extended power-law component is detected.

\section{Discussion}
\label{discussion}

We have developed a Comptonization model aimed to describe the prompt spectral emission 
of GRBs over a broad energy range. 
The assumption is the presence of two ingredients: a  thermal bath of soft X-ray photons 
and the existence of a hot plasma outflow  which  should be formed as a consequence of the explosion of a star. 
Both ingredients were also assumed by \citet{Lazzati00} for their Compton-drag model to interpret 
the prompt emission from GRBs associated with supernovae. 
In our model, the  hot plasma outflow   has a bulk velocity which is initially subrelativistic 
($V_{\rm b}/c \la 0.1-0.2$) and later it becomes relativistic outside the star photospheric radius. 
Note that a model consisting of an early subrelativistic  followed by a relativistic phase
has been proposed by \citep{Lyutikov03} in the context of GRBs originated by electromagnetic
outflow.

Our model consists of two parts: the first one describes the spectra up to  the peak energy in the EF(E) 
diagram and is obtained from the
solution of the Fokker-Planck expansion of the radiative transfer equation  which takes into account multiple scattering  of a BB-like seed photon population off a hot and thick
electron plasma [Compton cloud (CC)].  The relatively soft  BB population ($\ktbb \lax$ 1  keV)  presumably originates in the  relatively cold star photosphere and illuminates CC 
  (see Fig. \ref{geometry}).   
The spectral output of this early phase is characteristic of saturated Comptonization, but the shape slightly 
deviates from a pure Wien law because of the modification due to outflow motion 
(see Figs. \ref{plot_green_func} and \ref{conv_spectrum}). 
In this case, the peak of the spectrum in the EF(E) diagram falls slightly below $\Ep \sim 4 \kte$,
with respect to that for the standard Comptonization where the bulk outflow velocity $V_b=0$.

The second part of our model describes the prompt emission spectra beyond the peak energy  $E_p$. This was obtained by an assumption that, beyond the photosphere, the outflow velocity becomes relativistic and provides a further upscattering of the already Comptonized spectrum. We discuss now some relevant points related to our model.

It is important to estimate
the  thermalization optical depth $\tau_{\rm eff}$, which also appears
in the definition of $\delta$ (see Eq. \ref{delta_def}). This quantity indeed is not
an explicit parameter of the Comptonization equation (\ref{energy_op_full}).
 Using the best-fit values of $\kte$ and $\delta$ for GRB990705
(with $\beta_0\lax0.2$) we find $\tau_{\rm eff} \sim 2$. 

From the continuity equation in spherical symmetry we find
\begin{equation}
 N_{\rm e} \approx 4 \times 10^{14}  \frac{(\dot{m}/10^{-5})}{(\xi_{\Omega}/0.5) r_{14}^2 (\beta_0/0.1)} ~~{\rm cm^{-3}},
\label{cont_eq_b}
\end{equation}
where $\dot{m}\equiv \dot{M}/\dot{M_{\odot}}$ ($\dot{M_{\odot}} \approx 10^{33}$ g s$^{-1}$), $r_{14} \equiv R/(10^{14} \rm{cm}$),
$\xi_{\Omega}=2\pi (1-{\rm cos}~\theta$) and $\theta$ is the half opening angle of the subrelativistic outflow.

Then optical depth of corona located on the top of the star is
\begin{equation}
\tau_0=N_{\rm e}\sigma_{\rm T}H\sim7~\frac{(\dot m/10^{-5})(H/3\times10^{10} ~\rm{cm})}{(\xi_{\Omega}/0.5) r_{14}^2(\beta_0/0.1)},
\label{tau_corona}
\end{equation}
where $H$ is the thickness of the hot corona.
\subsection{The resulting GRB luminosity}
\label{GRB_luminosity}

The total luminosity of the observable GRB emission can be estimated   analytically
using the same procedure defined in ST80 and CT95. Let us first define
the BB luminosity illuminating the hot Compton cloud  as 
\begin{equation}
L_{bb,il}= 5\times10^{46}(\xi_{\Omega}/0.5)(kT_{\rm bb}/0.1~{\rm keV})^4r^2_{14}~{\rm erg}~{\rm s^{-1}}
\label{illum_BB_luminosty}
\end{equation}
Using the formula  for the  enhancement factor  $\eta_{\rm comp}$ (see equation \ref{enhancement_factor}) for the 
case of saturated Comptonization ($\alpha\ll 1$) we obtain 

\begin{equation}
L^{\rm comp}_{\rm grb}=\eta_{\rm comp}L_{bb,il}\sim 5\times10^{49}(T^{\rm (e)}_{100}/T^{\rm (bb)}_{0.1})(\xi_{\Omega}/0.5)(T^{\rm (bb)}_{0.1})^4r^2_{14}~{\rm erg}~{\rm s^{-1}},
\label{L_grb}
\end{equation} 
where $T^{\rm (e)}_{100} \equiv \kte/(100~{\rm keV})$ and $T^{\rm (bb)}_{0.1} \equiv \ktbb/(0.1~{\rm keV})$, respectively.

If the high-energy power law index of the hard component, which is presumably formed due to the inverse Compton effect, is $\alpha_{\rm b}\gax 2$ then  the related enhancement factor $\eta_{\rm rel}$
\begin{equation}
\eta_{\rm rel}=\frac{\alpha_{\rm b}(\alpha_{\rm b}+3)}{(\alpha_{\rm b}-1)(\alpha_{\rm b}+4)}
\label{alpha_b}
\end{equation} 
 is order of 1.
Thus the resulting  enhancement factor $ \eta_{res}\sim \eta_{\rm comp}$ and the GRB resulting luminosity 
$L_{\rm grb}=\eta_{\rm rel}L^{\rm comp}_{\rm grb}\sim L^{\rm comp}_{\rm grb}$.

\subsection{On the origin of the high-energy power-law component}

As we already outlined in Section \ref{sect_powerlaw}, subrelativistic bulk motion Comptonization 
is characterized by a rollover feature around an energy which is of the order of
the electron temperature, slightly modified by the presence of matter velocity field.
In any case, this process cannot give rise to any PL component extending up to GeV as observed in GRBs
\citep[e.g.,][]{abdo09}.
Actually a second component must be included in the X/$\gamma$-ray spectral model. If the
subrelativistic phase Comptonization spectrum and the high-energy PL have  different origin, 
then in terms of spectral modeling  the latter could be described by simply adding a PL 
spectral feature.
However, a physically unconnected origin of the two components seems  unlikely, 
given that all the  GRB spectra show  a smooth joining of the low-energy and high-energy 
features around the  peak energy, and the GeV emission  for most {\it Fermi} GRBs is 
consistent with the extrapolation of the MeV component \citep[][hereafter Z11]{zhang11}.

Here, we suggest this physical link in terms of the solution of Equation (\ref{up_scattering}) 
where the bulk Comptonization spectrum of the subrelativistic phase
(Eq. \ref{energy_op_full}) is convolved with a broken-PL GF (Eq. \ref{green_func_up}) leading to the emergent
broad-band spectrum (see Figs.  \ref{spec_parameters}-\ref{efe_grbcomp_band}).

The critical point to be addressed is of course the nature of the upscattering process that gives rise to the suggested  GF (Eq. \ref{green_func_up}).
The apparently non-thermal origin of the high-energy GRB component leads to include mechanisms of
particle non-thermal acceleration in shock regions. 
The standard fireball model actually allows the possibility to consider shocks as possible sites 
for particle acceleration, which eventually lead to non-thermal distributions $N_{\rm e}(E) \propto E^{\rm -p}$.
%
In this context, the high-energy PL would be produced through synchrotron and/or inverse 
 processes by this non-thermal particle bath. An intermediate case where the electron population  
 has a Maxwellian distribution  plus a supra-thermal tail was also considered by \cite{tavani96}.

One of the key assumptions of the fireball scenario is that the GRB outflow is highly relativistic
since the beginning of the event, with $\Gamma \ga 100$. In our model, however, the first phase of 
the GRB phenomenon is subrelativistic. How to conciliate
this significant difference? Is really an unavoidable requirement that the photosphere 
is ultra-relativistic?
Let us consider this possible scenario: when the top bounce of the hot shell emerges
from the star photosphere with bulk velocity $V_{\rm b}/c \la 0.1$, the underlying star
photospheric luminosity is huge ($L_{*} \ga 10^{10} L_{\rm Edd}$), and it illuminates
the shell itself, which subtends a solid angle $\Omega/4\pi <1$.
 This particular configuration (hot plasma, anisotropic high radiation pressure) could be suitable
to give rise to an efficient plasma acceleration through the so-called ``Compton-rocket effect'' \citep{cod81}.
For instance, in the case of an illuminating point source, the ratio of the radiative forces acting on a hot
and cold plasma, respectively, is $f_{\rm hot}/f_{\rm cold}$ = 2/3 $<(\gamma\beta)^2>$  \citep{odell81},
where $<(\gamma\beta)^2>$ is the quadratic average momentum of a Maxwellian electron distribution.
Numerical simulations by \cite{rh00} show that  electron-positron jet populations with 
non-thermal distribution illuminated by an underlying accretion disk can be accelerated up to Lorentz 
factors of several tens. 


It is beyond the scope of our paper to investigate quantitatively the hydrodynamical
configuration leading to plasma acceleration under
the underlying star radiation field.  Here we just claim a qualitative scenario where the accelerated plasma 
may interact with  the interstellar medium and/or star wind producing 
shocks with associated particle non-thermal populations. The high-energy photons of the subrelativistic 
phase then interact with these electrons via inverse Compton  effect (Fig. \ref{geometry})  and the GF of such a process in given by equation (\ref{green_func_up}).
Note that in this case, the radiative production mechanism is not qualitatively different from
the standard fireball model. No matter whether synchrotron or inverse Compton dominate the
spectral formation, what is important is the \emph{total} electron  Lorentz factor,
$\Gamma_{\rm tot} \sim \Gamma_{\rm sh} \gamma_{e}$, where $\Gamma_{\rm sh}$ and  $\gamma_{e}$
are the Lorentz factors of the expanding shell and of the electron in the shell frame, 
respectively.
A quantitative treatment of the radiation transfer problem is needed in this case.

Other models have been proposed to explain the GRB prompt emission in the framework of inverse
Compton processes \citep[see reviews by][]{m06,gehrels09}. 
The key difference between our  model and the other photospheric or synchrotron scenarios 
is that in our model  case \emph{the energy peak of the EF(E) diagram is independent of any Lorentz $\Gamma_L-$factor  associated to relativistic expansion.} The main parameter driving the peak position (the electron temperature $\kte$) is essentially the same in the fluid and observer frame (apart from cosmological redshift effects), 
{unlike that suggested by \citet{ghirl12}  who claim that the extension of $E_p$ energy 
of the $EF(E)-$diagram to  a few MeV is mostly due to distribution of Lorentz factors in  
the relativistic shell (or jet).} 

\subsection{The scenario in the light of the \fermi\ results}

The thermal nature of GRB spectra is still subject to debate. Blackbody components 
are predicted in various models \citep{mr00,Peer06,Thompson07,toma11} and actually they have been found 
in time resolved GRB spectra using   the {\it Beppo}SAX Gamma-Ray Burst Monitor and Wide Field Camera 2
\citep{frontera01} and 
BATSE  instruments \cite[][RP09]{ghirlanda03,ryde04,ryde05}.  However, from the
recent broad band observations with the \fermi\ GRBM plus LAT instruments it is not so clear whether 
the existence of a thermal component  is a general property of GRBs. 
Indeed, on the basis of the systematic and detailed time-integrated and time-resolved
analysis of a sample of 17 GRB data observed with GBM and LAT, Z11 found that most of the 
GRBs in this sample could be fitted with a Band  function over
the whole \fermi\ energy range. Only in two cases (GRB 090902B and possibly
GRB 090510) the spectrum was peculiar, in that it could be described by a
BB+PL model, similarly to what reported in RP09.
The paradigm of Z11 is that the Band-only spectra are better consistent with
a Poynting-dominated rather than baryon-dominated flow, thus discarding in
most (albeit not all) cases a photospheric scenario.
Against a pure photospheric emission the authors argue that  the extension is up
to GeV energies.

\begin{figure}[ptbptbptb]
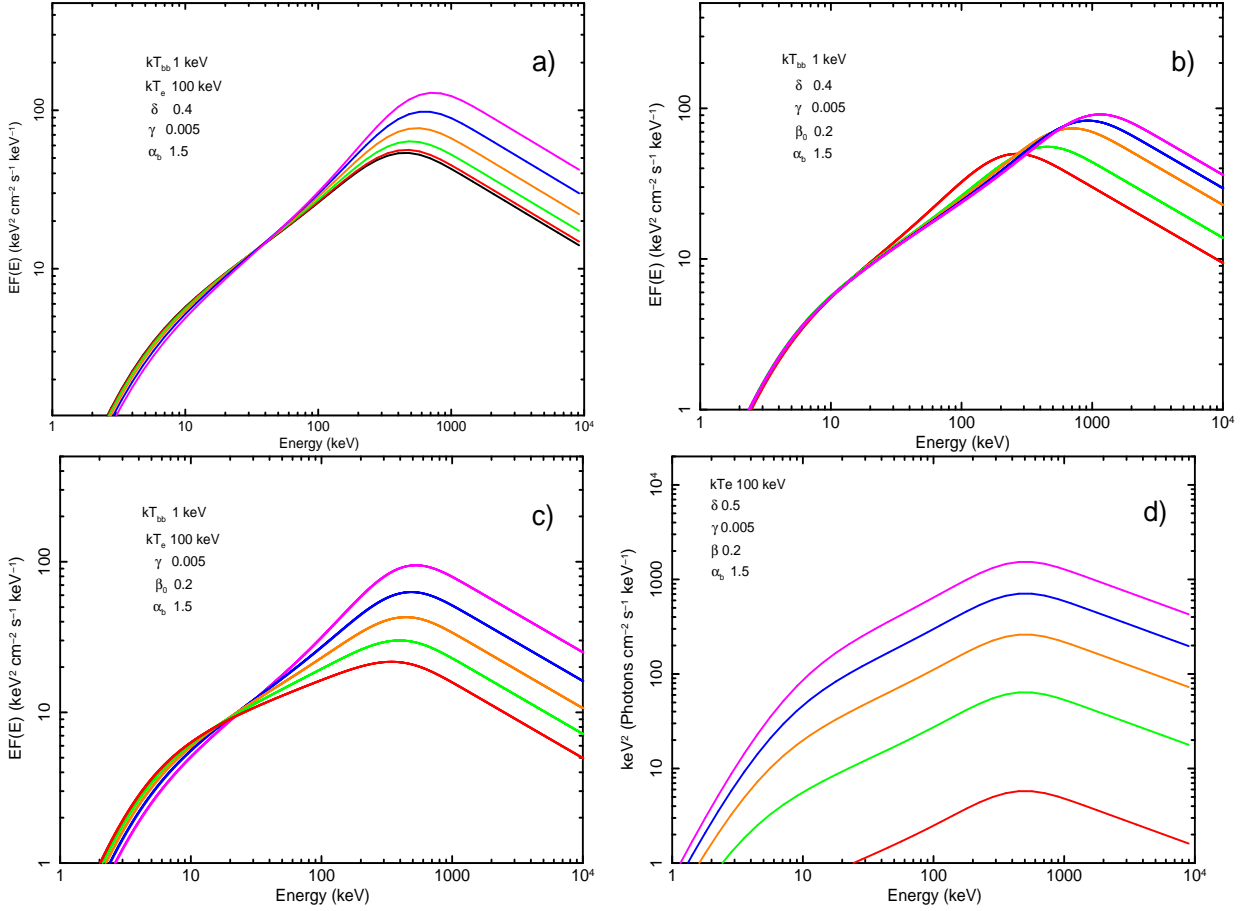

\includegraphics[width=6cm,height=8cm,angle=-90]{f6a.eps}
\includegraphics[width=6cm,height=8cm,angle=-90]{f6b.eps}
\includegraphics[width=6cm,height=8cm,angle=-90]{f6c.eps}
\includegraphics[width=6cm,height=8cm,angle=-90]{f6d.eps}
\caption{Simulated EF(E) spectra obtained as solutions of Eqs. (\ref{energy_op_full}) and (\ref{up_scattering}) for different sets of the Compton parameters values. In the top left of each panel, the fixed parameters are 
reported. The low-energy steepening of the spectra is the transition towards the Rayleigh-Jeans region of the 
seed BB spectrum. 
{\it Panel a)}: $\beta_0$=0 (black), 0.1 (red) , 0.2 (green), 0.3 (orange), 0.4 (blue), 0.5 (pink). 
{\it Panel b)}: $\kte$=50 keV(red), 100 keV (green),150 keV (orange), 200 keV (blue), 250 keV (pink). 
{\it Panel c)}: $\delta$= 0.8 (red), 0.7 (green), 0.6 (orange), 0.5 (blue), 0.4 (pink).
{\it Panel d)}: $\ktbb$=0.5 keV(red), 1 keV (green), 1.5 keV (orange), 2 keV (blue), 2.55 keV (pink).}
\label{spec_parameters}
\end{figure}

The preponderance of the non-detection of high temperature BB-like features in the \fermi\ GRB spectra
as reported in Z11 is not, in our opinion, necessarily in contrast with our
photospheric scenario.
Indeed, in our proposed model the seed thermal photons and
the electrons are decoupled and the low-energy slope of the emerging spectrum
(up to the peak energy $\Ep$) is mostly dictated by the optical depth and  plasma temperature of the hot corona 
and slightly  modified by subrelativistic outflow velocity $V_{\rm b}\ll c$
(see Figs. \ref{conv_spectrum} and \ref{spec_parameters}).
The Band function itself  can thus be yet the result of photospheric Comptonization
process, but with a different configuration of the radiation field
and electron plasma (decoupled) with respect to the standard scenario
(where both are coupled). It is also worth noting that the above-mentioned BB features
detected in BASTE and \fermi\ spectra have characteristic observer-frame
temperatures
of orders of hundreds keV, and thus must not be confused with
the cool BB seed photons ($\lax 1$ keV) of our model. 
More specifically, we identify these hot BB component with the optically thick subrelativistically expanding
plasma of the earlier phase in our scenario.

Actually, an important theoretical prediction of our model, to be tested with observations,
is the spectral steepening  at low-energies in comparison with the Band function
(see Fig. \ref{efe_grbcomp_band}). 
The change of slope is predicted to occur below $\sim 4 \ktbb$ keV in the EF(E) diagram,
where $\ktbb$ is the seed photon BB temperature, and it does represent the low-energy tail 
(Rayleigh-Jeans law) of the input BB. Thus a spectral slope $F(E) \propto E^2$ should
be a universal GRB observational feature at soft X--ray energies 
(see also Fig. \ref{spec_parameters}). Unfortunately, this issue cannot be solved
by \swift\ or \fermi\  as their energy thresholds are about 15 keV or 8 keV, respectively.
The \sax\ GRBs observed with both GRBM and WFCs are more suitable to be used to test
our model, even if the lower threshold  of the passband (2 keV) is nearby the limit of the energy
range foreseen for the seed photons. A test of our model  has been performed using the {\it BeppoSAX} data(Frontera et al.  2012,  in preparation). 

\section{Conclusions}
\label{conclusions}

We have developed a spectral model aimed to describe the broadband prompt emission of GRBs.
We propose that the spectral emission during the prompt phase, phenomenologically modeled by the
 Band  function, is the result of an earlier phase where soft BB-like photons are Comptonized by an 
optically thick and hot electron shell ($T_{\rm e} \sim 10^9$ K), something like a Compton cloud  sub-relativistically moving outwards the star surface. 
On the other hand in the relativistic phase, these Comptonized photons are subjected to a second upscattering process which can be mathematically described
by a broken power-law Green function whose spectral index models the high-energy slope of the 
Band function.
An important prediction  of our proposed model is that the peak energy in the EF(E) diagram originated in 
the early subrelativistic phase (see  Fig. \ref{geometry}) is directly related to the plasma temperature of the hot plasma $T_e$.   We demonstrate that {\it the resulting luminosity of $X/\gamma$-rays luminosity of GRB $L_{\rm grb}$ is proportional to $(\kte)^2$} (see Eq. \ref{L_subrel}).   

In fact , $L_{\rm grb}$ is a product of the Comptonization  enhancement factor $\eta_{\rm comp}$ and luminosity of the soft blackbody photons $L_{bb,il}$ but in the case of the saturated Comptonization, 
when $\alpha\ll1$,    $\eta_{\rm comp}\propto \kte$ (see Eq. \ref{enhancement_factor})  but $L_{\rm bb,il}$ (or $L_{\rm soft}$) is also proportional to  $\kte$ because the surface area of Compton cloud illuminated by the soft photons is proportional  to $\kte$ (see Eq. \ref{hot_spot_area}). 
Thus we claim that { \it the model dependence of  $L_{\rm grb}\propto (\kte)^2$ on the hot plasma temperature $T_{\rm e}$
explains the observed  Amati  relation  in which  $E_{iso}=\int L_{\rm grb}(t)dt \propto E_{\rm p}^2$}.  
It is worth noting that the peak energy $\Ep$ of the emergent Wien spectrum should be equal to $3\kte$. 

Our model also explains  both the  observed long and short temporal  variability during GRB prompt emission phase. 
The former one is related to time scale of the radiation formed in the sub-relativistically moving  Compton cloud while the short time scale is related to time scale of radiation formed in the sub-relativistically moving Compton hot spot while the short time scale is related to 
the time scale of radiation coming to the Earth observer from the relativistically driven outflow.

We do not quantitatively investigate the physical conditions underlying the origin of the high-energy component.
Different scenarios, not necessarily excluding each other, can be operating such as a second inverse Compton process  off non-thermal electrons accelerated  by super Eddington radiation pressure  (for which $L \gg L_{Edd}$)
as the hot shell moves out of  the star.
Self-synchrotron Compton or direct synchrotron effects can not be excluded.
An important prediction  of our  model is that the peak energy in the EF(E) diagram originates in 
the early sub-relativistic phase and is  proportional to plasma temperature  $\kte$ and the resulting luminosity 
$L_{\rm grb}$ is  proportional to $(\kte)^2$. 
This dependence   is  the same, after cosmological corrections,  in the source and observer frame. 
In fact, no fine tuning related to some Lorentz $\Gamma$-factor of relativistic 
expansion is required. We claim that this result of our model {\it naturally} explains the physical origin of the Amati relation.

A systematic application of our model to a sample of time-resolved GRB prompt spectra of the {\it Beppo}SAX archive
by Frontera et al. (2012, to be submitted) proves that in \emph{all the cases} the \chiq-values are equal or better than those  derived with the Band function. To our knowledge, this is the first time that a model for the GRB prompt emission other  than the Band function has been developed under the X-ray spectral fitting package 
XSPEC.

{\it Acknowledgements}
The authors acknowledge very productive discussions with Davide Lazzati, Pawan Kumar and Alexandre   Chekhtman which strongly improved the quality of our paper. Important suggestions have been given by the anonymous referee, which have allowed us to better clarify the main topics of our model.

\appendix
\section{Numerical solution of Fokker-Planck radiative transfer equation using finite differences}
\label{sweep}

We  consider the Fokker-Planck expansion of the radiative transfer equation for the zero-moment occupation number N(x), in the presence of subrelativistic
bulk motion (see Eq. \ref{energy_op_full}):
\begin{equation}
f_{\rm b}x^2\frac{d^2N(x)}{d^2x}+(x^2+4f_{\rm b} x+\delta x)\frac{dN(x)}{dx}+(4x-\gamma)N(x)=- \frac{\varphi(x)}{x^3},
\label{sweep_total}
\end{equation}
where $x \equiv E/\kte$, $\Theta \equiv \kte/\me c^2$, $f_{\rm b}=1+3\beta_0^2/\Theta$, while  $\gamma$ and $\delta$ are defined 
in Sect. \ref{sect_bulk_solution}. For $\varphi(x)$ we assume a BB spectrum (see Eq. \ref{bb_seed}).

In terms of the zero-moment intensity $J(x)=N(x)x^3$ the equation becomes
\begin{equation}
fx^2\frac{d^2J(x)}{d^2x}+(x^2-2fx+\delta x)\frac{dJ(x)}{dx}+(x-\gamma-3\delta)J(x)=- \varphi(x).
\end{equation}
In order to provide logarithmic binning of the a dimensional energy $x$, we introduce the
new variable $y$ such that $x=e^y$, thus obtaining the new equation in terms of this variable
\begin{equation}
f\frac{d^2J(y)}{d^2y}+(e^y-3f+\delta)\frac{dJ(y)}{dy}+(e^y-\gamma-3\delta)J(y)=- \varphi(y).
\end{equation}

For numerical integration, we divide the domain of $y$ from $y_{\rm min}$  to $y_{\rm max}$ 
into N equally-stepped sizes $h=(y_{\rm max} - y_{\rm min})/N$, so that equation  can be presented, using finite differences, as
 
\begin{equation}
p_{\rm n} \frac{J_{\rm n+1} -2 J_{\rm n} + J_{\rm n-1}}{h^2}+g_{\rm n} ~\frac{J_{\rm n+1} - J_{\rm n}}{h}+r_{\rm n} J_n=- \varphi_{\rm n},
\label{finite_diff}
\end{equation}
where 
\begin{equation}
p_{\rm n}=f, ~~~~~~~ g_{\rm n}=(e^{y_{\rm n}} -3f-\delta)~~~~~~{\rm and} ~~~~~~~~r_{\rm n}=(e^{y_{\rm n}}+3\delta - \gamma).
\end{equation}

The two boundary conditions for the intensity J(x) are
\begin{equation}
J[0]=0, ~~~~J[N]=0,
\label{bc_intensity}
\end{equation}
resembling the physical conditions $J(x)=0$ for $x \rightarrow 0$ and  $x \rightarrow \infty$.

Collecting the terms with the same index in equation (\ref{finite_diff}) and using the boundary conditions
(\ref{bc_intensity}), we may rewrite the equation as

\begin{equation}
a_{\rm n}J_{\rm n-1}+b_{\rm n}J_n+c_{\rm n}J_{\rm n+1}=g_{\rm n},
\label{num_eq}
\end{equation}
with coefficients $a_{\rm n}$, $b_{\rm n}$, $c_{\rm n}$ defined as

\begin{eqnarray}
\cases{{\displaystyle a_0=0,~~~~b_0=-1,~~~~~~~c_0=0,~~~~~\varphi_0=0,  ~~~~~~~~~~~~~~n=0}
\cr
{\displaystyle a_{\rm n}=\frac{p_{\rm n}}{h^2},~~b_{\rm n}=-\left(\frac{2p_{\rm n}}{h^2}+\frac{g_{\rm n}}{h}-r_{\rm n}\right), ~~~~c_{\rm n}=\frac{p_{\rm n}}{h^2}+\frac{g_{\rm n}}{h}, ~~\varphi_n=\varphi(t_n),~~~{\rm for}~~n=1,...N-1,}
\cr
{\displaystyle a_{\rm N}=0,~b_{\rm N}=1,~~c_{\rm N}=0, ~~~\varphi_{\rm N}=0, ~~~~~~~~~~n=N.}
}
\end{eqnarray}

For $n=0$ Equation (\ref{num_eq}) can be presented as

\begin{eqnarray}
J_0=L_{0}J_1+K_{0},~~~{\rm where}~L_{0}=-c_0/b_0, ~~{\rm and} ~~K_{0}=0.
\label{n_0_L_K}
\end{eqnarray}

Excluding $J_0$ from equation (\ref{num_eq}) for $n=1$, we obtain
\begin{eqnarray}
(a_1L_{0}+b_1)J_1+c_1J_2=g_1-a_1K_{0}.
\label{eps_1_eps_2}
\end{eqnarray}

Because $a_1L_{0}+b_1\neq0$ we also obtain that

\begin{eqnarray}
J_1=L_{1}J_2+K_{1},
\label{eps_1_eps_2_mm}
\end{eqnarray}
where we have introduced new coefficients

\begin{eqnarray}
L_{1}=-\frac{c_1}{a_1L_{0}+b_1}, ~~ {\rm and}~~K_{1}=\frac{g_1-a_1K_{0}}{a_1L_{0}+b_1}.
\label{L2_K2}
\end{eqnarray}

Continuing the process of exclusion of $J_n$ we obtain in general

\begin{eqnarray}
J_{\rm n}=L_{\rm n}J_{\rm n+1}+K_{\rm n},
\label{epsn_epsn+1}
\end{eqnarray}

\begin{eqnarray}
{\rm where}~~L_{\rm n}=-\frac{c_{\rm n}}{a_{\rm n}L_{\rm n-1}+b_{\rm n}},~~ {\rm and}~~K_{\rm n}=\frac{g_{\rm n}-a_{\rm n}K_{\rm n-1}}{a_{\rm n}L_{\rm n-1}+b_{\rm n}}.
\label{Ln_Kn}
\end{eqnarray}

Now, using the right boundary condition J[N]=0 and relation (\ref{epsn_epsn+1}) it is possible to find
\begin{equation}
J_{\rm N-1}=L_{\rm N-1}J_{\rm N}+K_{\rm N-1},
\end{equation}
and so on until all $J_{\rm N-2},~J_{\rm N-3},...J_{3},~J_{2}, ~J_{1}, ~J_{0}$ are calculated.
This algorithm was first introduced by Israel Gelfand is called by the sweep (progonka) method.


In order to reproduce the unbroken PL Component of the high-energy spectrum, we solved Equation 
(\ref{up_scattering}) for the zero moment intensity, replacing $\varphi(x)$ with $F_{\rm tb}(x)$ which gives

\begin{equation}
x^2\frac{d^2J(x)}{d^2x}-2x\frac{dJ(x)}{dx}- \alpha_{\rm b} (\alpha_{\rm b}+3) J(x)=-F_{\rm tb}(x).
\label{sweep_up}
\end{equation}

As we have already shown in Sect. \ref{sect_powerlaw}, this is equivalent to calculation of
the convolution integral in Equation (\ref{convolution}) with the GF given by
formula (\ref{green_func_up}) and $\varphi(x)=F_{\rm tb}(x)$.
The procedure for the numerical solution of equation (\ref{sweep_up}) is identical to that
used for solving Equation (\ref{sweep_total}) but  just changing the expression
for the cofficients $a_{\rm n}, ~b_{\rm n}$ and $c_{\rm n}$.

\section{Development of a spectral model for XSPEC}
\label{xspec_model}

We  develop a numerical model for the X-ray spectral fitting package XSPEC.
The model can be actually divided into two parts, one strictly physical, in terms of the 
parameter setting, and the second one which is more phenomenological.

\vskip 0.25cm
\emph{Step~1: Comptonization in subrelativistic outflow}
\vskip 0.25cm

The first part of the model computes the spectrum obtained as a result of Comptonization
of a BB-like seed photon population off a hot electron corona which is subrelativistically expanding,
according to equation (\ref{fp_equation}). For practical purposes, actually we solved equation (\ref{energy_op_full}),
which is obtained from the variable separation method, with the bulk parameter $\delta$ defined in Eq. (\ref{delta_def}).

The input seed spectrum is given by the Planck's law
\begin{equation}
\varphi(x)= N_{\rm bb}( \kte)^3  \frac{x^3}{e^{(\kte/\ktbb)x}-1},
\label{bb_seed}
\end{equation}
where $x \equiv E/\kte$, while $\ktbb$ and $\kte$ are the BB and electron temperatures, respectively.

The normalization constant $N_{\rm bb} \equiv R^2_{9}/D^2_{\rm Mpc}$, where $R_{9}$ is the BB apparent photospheric radius
in units of $10^{9}$ cm  and $D_{\rm Mpc}$ is the source distance in Mpc, respectively.
Note that this is the source term appearing in the right-hand side of equation  (\ref{energy_op_full}). 
We numerically solved the latter equation using a fast finite-difference method (see Appendix \ref{sweep})
rather then performing a convolution with  GF defined in Eq. (\ref{convolution}). 
The free-parameters of the model at this step are  the BB temperature and normalization 
($\ktbb$ and $N_{\rm bb}$, respectively), the electron plasma temperature $\kte$, the bulk parameter $\delta$, 
the $\gamma$-parameter (which, in fact,  is the inverse of the Comptonization 
parameter $Y$) and the second power of the (constant) outflow velocity $\beta_0^2$.

We note that the outflow velocity $\beta_0$ appears in Equation  (\ref{energy_op_full}) both explicitly in the term $\fb=1+\beta_0^2/(3\Theta)$
and implicitly  in the definition of the $\delta$-parameter (see Eq. \ref{delta_def}). The linear dependence on  $\beta_0$  via the bulk parameter 
$\delta$ is related to  the effect of Fermi first-order photon down (up)-scattering
and the net result is photon red-shift or blue-shift, depending on the sign of ${\bf \nabla} \cdot {\bf \Vb}$,
while the quadratic dependence of $\beta_0$ in $\fb$ provides the second-order Fermi process and represents
an up-scattering contribution (when averaged over angles) term in addition to that
coming from the electron thermal velocity component.
By setting $\beta_0^2\ll 3\Theta$, it is possible to find the analytical solution for the emerging spectrum $F_{\rm tb}(x)$ using 
convolution (\ref{convolution}) and  the analytical presentation of  the GF
(see Eq. \ref{green_func}).
A comparison between the numerical and analytical solution in this limit has 
shown an excellent agreement between the two cases.

\vskip 0.25cm
\emph{Step 2: high-energy powerlaw component}
\vskip 0.25cm

\begin{figure}
\includegraphics[width=12cm,height=14cm,angle=-90]{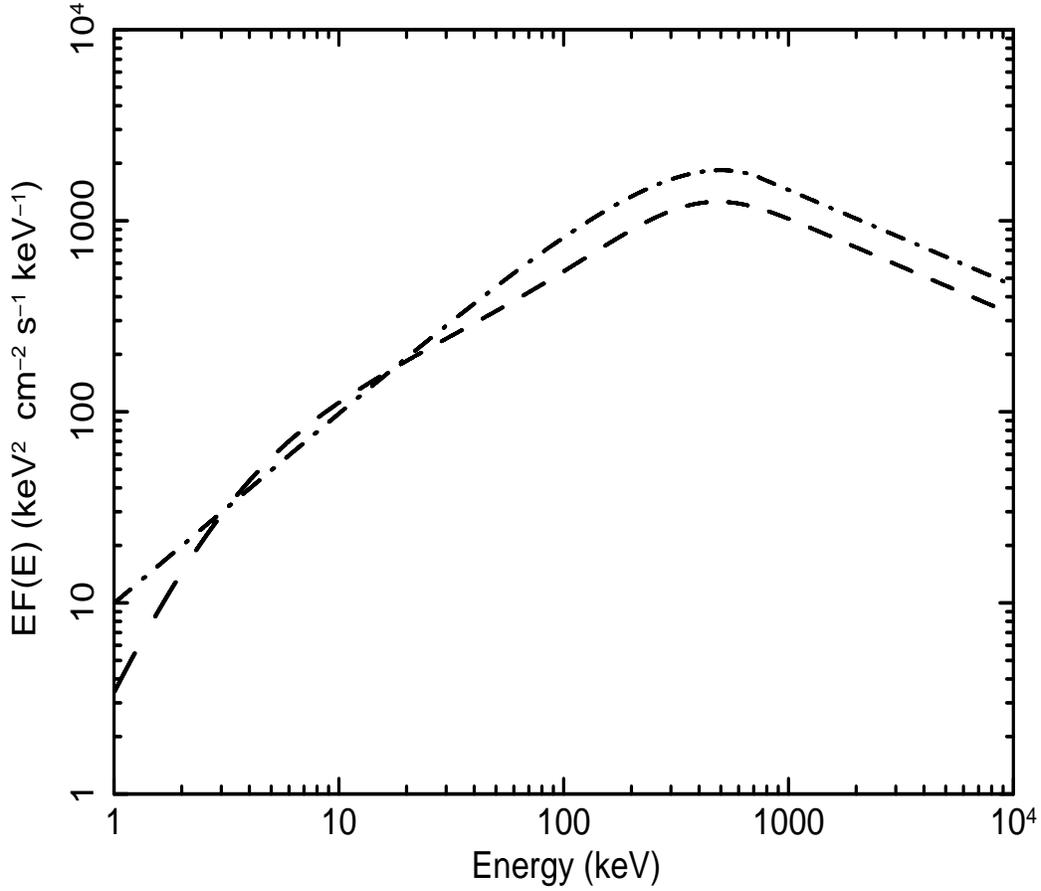}
\caption{Comparison of the EF(E) spectra in the 1 keV -- 10 MeV energy range of our Comptonization model ({\it dashed line}) 
obtained from the solution of equations (\ref{energy_op_full}) and (\ref{up_scattering}) and the Band 
function ({\it dotted-dashed line}).
The parameters of the simulation are: $\Gamma_1=-1.5$, $\Gamma_2=-2.5$ and $E_0$=350 keV for the Band function, $\ktbb$=1 keV, $\kte$=100 keV, 
$\delta$=0.5,  $\gamma=5 \times 10^{-3}$, $\beta_0=0.4$  and $\alpha_{b}$=1.5, for the Comptonization spectrum.}
\label{efe_grbcomp_band}
\end{figure}

We should  point out that in principle a \emph{general} broken-PL GF is described
by two indexes for $E < E_0$ and $E > E_0$, while in the GF of equation (\ref{green_func_up}) they both
depend on $\alpha_{\rm b}$. However, for our purposes, it is important that the red wing of the GF, $\propto E^{\alpha_{\rm b}+3}$,
is significantly steeper than the blue wing $\propto E^{-\alpha_{\rm b}}$. Actually, 
it is this up-scattering term
of the GF which dominates the convolution process.
 We show an example of  spectrum obtained from the solution of
 Eqs.  (\ref{energy_op_full}) and (\ref{up_scattering}) compared with the Band function in 
 Figure \ref{efe_grbcomp_band}.
It is evident their similarity, but we also draw the attention of the reader to the change
in the slope at low energy of our model with respect to the Band function. The low-energy steepening occurs at  $E \la 4~\ktbb$
 in a EF(E) diagram, where $\ktbb$ is the temperature of the seed BB photons, and it is related to the Rayleigh-Jeans
part of the Planckian distribution, where $F(E) \propto E^2$. For the case reported in Fig. \ref{efe_grbcomp_band} we assumed 
$\ktbb=1$ keV, so that the steepening occurs below 4 keV.
We subsequently implemented our model in the \xspec\ package and tested it on GRB990705 (z=0.842; Amati 2006), observed by the  BeppoSAX/WFC+GRBM instruments. 



\clearpage
\begin{thebibliography}{}
\bibitem[Abdo  et al. (1992)]{abdo09}
Abdo, A. A. et al.  2009, Science,  323, 1688 
\bibitem[Amati (2006)]{amati06}
Amati, L.   2006, MNRAS,  372, 233 
\bibitem[Amati  et al. (2009)]{amati09}
Amati L., Frontera, F. \& Guidorzi, C.  2009, \aap,  508, 173 
\bibitem[Amati  et al. (2002)]{amati02}
Amati, L. et al.   2002, \aap,  390, 81 
\bibitem[Amati  et al. (2008)]{amati08}
Amati L. et al.   2008, MNRAS,  391, 577 
\bibitem[Band et al. (1993)]{band93}
Band D. et al.   1993, \apj,  413, 281
\bibitem[Becker et al. (2007)]{bw07}
Becker, P.A. \& Wolff, M.T.   2007, \apj,  654, 435 
\bibitem[Beloborodov (2010)]{beloborodov10}
Beloborodov, A.M.   2010, MNRAS,  407, 1033 
\bibitem[Blandford \& Payne (1981)]{bp81a}
Blandford R.D. \& Payne, D.G.  1981, MNRAS,  194, 1041 (BP81)
\bibitem[Bychkov et al.  (2006)]{b06}
Bychkov, V., Popov, M.V.,  Oparin, A.M., Stenflo, L. \& 
Chechetkin, V.M. 2006, Astronomy Reports,  50, 298 
\bibitem[Chakrabarti \&  Titarchuk  (1995)]{ct95}
Chakrabarti, S. K. \& Titarchuk, L. G. 1995, \apj,  455, 623
\bibitem[Chardonnet et al. (2010)]{cct10}
Chardonnet, P., Chechetkin, V. \& Titarchuk, L. 2010, \apss,  325, 153 
\bibitem[Cheng  \& Odell (1981)]{cod81}
Cheng, A.Y.S. \& Odell, S.L. 1981, \apjl,  251, L49
\bibitem[Colpi  (1988)]{colpi88}
Colpi, M. 1988, \apj,  326, 223
\bibitem[Cowsik \&  Lee (1982)]{cl82}
Cowsik, R. \& Lee, M.A. 1982, Royal Society of London Proceedings Series A,   383, 409
\bibitem[Crider  et  al. (1998 )]{crider98}
Crider, A.,  Liang, E.P. \& Preece, R.D.1998,  in American Institute of Physics Conference Series, Gamma-Ray Bursts,  Vol. 428, 4th Hunstville Symposium,   Eds. C.A.Meegan, R.D. Preece, \& T.M. Koshut,  428, 359-363
\bibitem[Crider  et  al. (1997 )]{crider97}
Crider, A. et al.  1997,  \apjl,   479, L39
\bibitem[Farinelli  et  al. (2008 )]{f08}
Farinelli, R., Titarchuk, L. Paizis, A. \& Frontera, F.   2008,  \apj,   680, 602, (F08)
\bibitem[Fiore  et  al. (2007 )]{fiore07}
Fiore, F., Guetta, D., Piranomonte, S.,  D'Elia, V.,  
Antonelli, L.A.  2007,  \aap,   470, 515
\bibitem[Frontera et al. (2000)]{frontera00}
Frontera, F. et al. 2000, \apjs, 127, 59
\bibitem[Frontera et al. (2001)]{frontera01}
Frontera, F. et al. 2001, \apjl, 550, L47
\bibitem[Fryxell et al. (2000)]{fryxell00}
Fryxell, B. et al. 2000, \apjs, 131, 273
\bibitem[Gal-Yam et al. (2009)]{gy09}
Gal-Yam, A. et al. 2009, \nat, 462, 624
\bibitem[Gao et al. (2012)]{Gao11}
Gao, H. Zhang, B.B. \& Zhang, B.  2012, \apj, 748, 134
\bibitem[Gehrels et al. (2012)]{gehrels09}
Gehrels, N., Ramirez-Ruiz, E. \& Fox, D.B. 2009, \araa, 47, 567
\bibitem[Ghirlanda et al. (2007)]{Ghirlanda07}
Ghirlanda, G., Bosnjak, Z., Ghisellini, G., Tavecchio, F. \& 
Firmani, C.  2007, \mnras, 379, 73
\bibitem[Ghirlanda et al. (2003)]{ghirlanda03}
Ghirlanda, G., Celotti,  A. \&  Ghisellini, G. 2003, \aap, 406, 879
\bibitem[Ghirlanda et al. (2012)]{ghirl12}
Ghirlanda, G. et al. 2012, \mnras, online Early, arXiv:1203.0003
\bibitem[Ghirlanda et al. (2010)]{Ghirlanda10}
Ghirlanda, G., Nava, L. \&  Ghisellini, G. 2010, \aap, 511, A43
\bibitem[Imshennik et al. (1999)]{imshennik99}
Imshennik, V.S., Kal'Yanova, N.L., Koldoba, A.V. \& 
Chechetkin, V.M. 1999, Astronomy Letters, 25, 206
\bibitem[Jakobsson et al. (2006)]{jakobsson06}
Jakobsson, P.  et al. 2006, \aap, 447, 897
\bibitem[Kelly et al. (2008)]{kkp08}
Kelly, P.L., Kirshner, R.P. \& Pahre, M. 2008, \apj, 687, 1201
\bibitem[Langer et al. (2007)]{Langer07}
Langer, N.,  Norman, C.A., de Koter, A., Vink, J.S., 
Cantiello, M. \& Yoon, S.-C. 2007, \aap, 475, L19
\bibitem[Laurent \& Titarchuk (2007)]{lt07}
Laurent, P. \& Titarchuk, L. 2007, \apj, 656, 1056 (LT07)
\bibitem[Laurent \& Titarchuk (1999)]{lt99}
Laurent, P. \& Titarchuk, L. 1999, \apj, 511, 289
\bibitem[Lazzati et al.  (2000)]{Lazzati00}
Lazzati, D., Ghisellini, G., Celotti, A. \&  Rees, M.J. 2000, \apjl, 529, L17
\bibitem[Lazzati et al.  (2009)]{lazzati09}
Lazzati, D., Morsony, B.J. \& Begelman, M.C. 2009, \apjl, 700, L47
\bibitem[Liang et al.  (2006)]{liang06}
Liang, E. et al.  2006, \apj, 646, 351
\bibitem[Lyutikov \& Blandford  (2003)]{Lyutikov03}
Lyutikov, M. \& Blandford, R.  2003,  arXiv:astro-ph/03035410
\bibitem[Lyutikov et al. (2003)]{lb03}
Lyutikov, M., Pariev, V.I. \& Blandford, R.  2003, \apj, 597, 998
\bibitem[Meegan et al. (1992)]{meegan92}
Meegan, C.A.  et al.   1992, \nat, 355, 143
\bibitem[M{\'e}sz{\'a}ros (2006)]{m06}
M{\'e}sz{\'a}ros, P.   2006, Reports on Progress in Physics,  69, 2259
\bibitem[M{\'e}sz{\'a}ros \& Rees (2000)]{mr00}
M{\'e}sz{\'a}ros, P.  \&  2000, \apj,  530, 292
\bibitem[Odell (1981)]{odell81}
Odell, S.L.  1981, \apjl,  243, L147    Thompson07
\bibitem[Peer et al.  (2006)]{Peer06}
Pe'er, A., M{\'e}sz{\'a}ros, P. \& Rees, M.J. 2006, \apj,  642, 995
\bibitem[Piran et al.  (1999)]{piran99}
Piran, T. 1999, \physrep,  314, 575
\bibitem[Raskin et al. (2008)]{Raskin08}
Raskin, C., Scannapieco, E.,  Rhoads, J. \& Della Valle, M. 2008, \apj, 689, 358
\bibitem[Rees \&  M{\'e}sz{\'a}ros  (2005)]{rm05}
 Rees, M. \&  M{\'e}sz{\'a}ros, P.  2005, \apj, 628, 847
 \bibitem[Renaud \& Henri  (2000)]{rh00}
 Renaud, N. \& Henri, G.  2000, Nuclear Physics B Proceedings Supplements,  80, C125
 \bibitem[Rybicki \& Lightman  (1979)]{rl79}
Rybicki, G.B. \& Lightman, A.P.  1979, Radiative Processes in Astrophysics, 
New York, Wiley-Interscience,  393 p.
 \bibitem[Ryde   (2005)]{ryde05}
Ryde, F.  2005, \apjl, 625, L95 
\bibitem[Ryde   (2004)]{ryde04}
Ryde, F.  2004, \apj, 614, 827
\bibitem[Ryde  \& Pe'er  (2009)]{rp09}
Ryde, F.  \& Pe'er, A.  2009, \apj, 702, 1211
\bibitem[Sunyaev \& Titarchuk (1985)]{st85}
Sunyaev, R.A.  \& Titarchuk, L.G.  1985, \aap,  143, 374 
\bibitem[Sunyaev \& Titarchuk (1980)]{st80}
Sunyaev, R.A.  \& Titarchuk, L.G.  1985, \aap,  86, 121 (ST80)
\bibitem[Tavani  (1996)]{tavani96}
Tavani, M.  1996, \apj,  466, 768 
\bibitem[Thompson (1994)]{thompson94}
Thompson, C.  1994, \mnras,  270, 480 
\bibitem[Thompson  et al. (2007)]{Thompson07}
Thompson, C., M{\'e}sz{\'a}ros, P. \& Rees, M.J.  2007, \apj,  666, 1012 
\bibitem[Titarchuk   et al. (2003)]{tkb03}
Titarchuk, L.  Kazanas, D.  \& Becker, P.A.  2003, \apj,  598, 411 
\bibitem[Titarchuk   et al. (1997)]{tmk97}
Titarchuk, L., Mastichiadis, A. \& Kylafis, N.D.  1997, \apj,  487, 834 (TMK97)
\bibitem[Toma  et al. (2011)]{toma11}
Toma, K., Wu, X.-F. \& M{\'e}sz{\'a}ros, P. 2011, \mnras,  415, 1663 
\bibitem[Vetere et al. (2006)]{vetere06}
Vetere, L.,  Massaro, E., Costa, E.,  Soffitta, P.  \&  Ventura, G. 2007, \aap,  447, 499 
\bibitem[Woosley  (2011)]{woosley11}
Woosley,  S.E. 2011, arXiv:1105.4193 
\bibitem[Woosley  et al. (2007)]{Woosley07}
Woosley,  S.E., Blinnikov, S. \& Heger, A.  2007, \nat, 450, 390 
\bibitem[Zhang  et al. (2011)]{zhang11}
Zhang, B.B.  2011, \apj, 730, 141, (Z11)
\bibitem[Zhang  \& Yan (2011)]{zy11}
Zhang, B.B. \& Yan, H. 2011, \apj, 726, 90
\end{thebibliography}

\begin{figure}
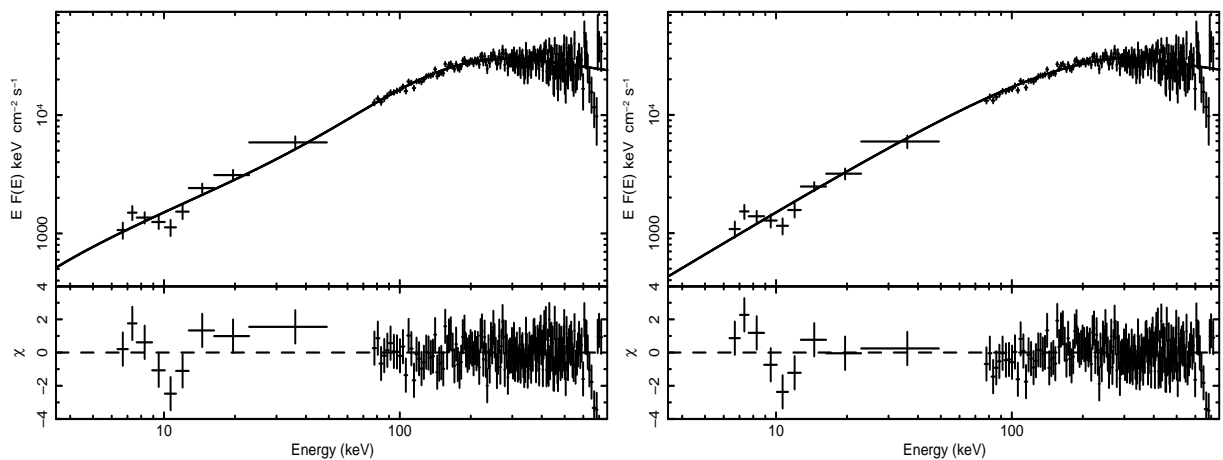

\includegraphics[width=6cm,height=8cm,angle=-90]{f8a.eps}
\includegraphics[width=6cm,height=8cm,angle=-90]{f8b.eps}
\caption{Unabsorbed EF(E) BeppoSAX/WFC+GRBM spectra of the prompt emission of GRB990705 (z=0.842; Amati 2006) and residuals
between the data and the model in units of $\sigma$ using our photospheric model ({\it left panel}) and the Band function
({\it right panel}). Best-fit parameters for both cases are reported in Section \ref{Amati_relation}.}
\label{grb990705}
\end{figure}


\end{document}